\documentclass{ws-procs9x6}

\newcommand{\lsim}{
\mathrel{\hbox{\rlap{\hbox{\lower4pt\hbox{$\sim$}}}\hbox{$<$}}}}

\newcommand{\gsim}{
\mathrel{\hbox{\rlap{\hbox{\lower4pt\hbox{$\sim$}}}\hbox{$>$}}}}

\begin{document}


\thispagestyle{empty}

\begin{flushright}
CERN-PH-TH/2005-078\\
hep-ph/0505018
\end{flushright}

\vspace{1.0truecm}
\begin{center}
\boldmath
\large\bf New Physics in $B$ and $K$ Decays
\unboldmath
\end{center}

\vspace{0.9truecm}
\begin{center}
Robert Fleischer\\[0.1cm]
{\sl CERN, Department of Physics, Theory Unit\\
CH-1211 Geneva 23, Switzerland}
\end{center}

\vspace{0.9truecm}

\begin{center}
{\bf Abstract}
\end{center}

{\small
\vspace{0.2cm}\noindent
Flavour physics offers interesting probes for the exploration of the Standard 
Model and the search for new physics. In these lectures, we focus
on $B$- and $K$-meson decays, introduce the concept of low-energy
effective Hamiltonians to describe them theoretically, and discuss how 
physics beyond the Standard Model may generically affect the roadmap 
of quark-flavour physics. We address then both the implications of the 
$B$-factory data for the $B_d\to J/\psi K_{\rm S}$ channel and the prospects 
of $B_s\to J/\psi \phi$ modes for hadron colliders, and discuss how the 
Standard Model may be challenged through $B_d\to \phi K_{\rm S}$. 
Finally, as an example of a systematic flavour strategy to search for new physics, 
we analyze puzzling patterns in the $B\to\pi\pi, \pi K$ data and study their 
interplay with rare $K$ and $B$ decays. 
}

\vspace{0.9truecm}

\begin{center}
{\sl Invited lectures given at the Lake Louise Winter Institute:\\
``Fundamental Interactions''\\
Chateau Lake Louise, Alberta, Canada, 20--26 February 2005\\
To appear in the Proceedings (World Scientific)}
\end{center}

\vfill
\noindent
CERN-PH-TH/2005-078\\
May 2005

\newpage
\thispagestyle{empty}
\vbox{}
\newpage
 
\setcounter{page}{1}


\title{New Physics in $B$ and $K$ Decays}

\author{R. FLEISCHER}

\address{CERN, Department of Physics, Theory Unit,\\
CH-1211 Geneva 23, Switzerland\\
E-mail: Robert.Fleischer@cern.ch}

\maketitle

\abstracts{
Flavour physics offers interesting probes for the exploration of the Standard 
Model and the search for new physics. In these lectures, we focus
on $B$- and $K$-meson decays, introduce the concept of low-energy
effective Hamiltonians to describe them theoretically, and discuss how 
physics beyond the Standard Model may generically affect the roadmap 
of quark-flavour physics. We address then both the implications of the 
$B$-factory data for the $B_d\to J/\psi K_{\rm S}$ channel and the prospects 
of $B_s\to J/\psi \phi$ modes for hadron colliders, and discuss how the 
Standard Model may be challenged through $B_d\to \phi K_{\rm S}$. 
Finally, as an example of a systematic flavour strategy to search for new physics, 
we analyze puzzling patterns in the $B\to\pi\pi, \pi K$ data and study their 
interplay with rare $K$ and $B$ decays. 
}

\section{Introduction}\label{sec:intro}
In flavour physics, the parity and charge-conjugation operators $\hat P$ and
$\hat C$, which describe the space-inversion operation and the replacement 
of all particles by their antiparticles, respectively, play a key r\^ole. After
the discovery that weak interactions are not invariant under parity and
charge-conjugation transformations in 1957, it was believed that the product 
of $\hat C$ and $\hat P$ was actually preserved. It came then as a big
surprise in 1964,\cite{CP-obs} when it was observed through the detection of 
$K_{\rm L}\to \pi^+\pi^-$ decays that this is actually {\it not} the case! 
The corresponding phenomenon is referred to as {\it CP violation}, and 
is a central aspect of flavour physics. The manifestation of CP violation
discovered in 1964 is ``indirect'' CP violation, which is described by a
complex quantitiy $\varepsilon_K$ and originates from the fact that the 
mass eigenstates of the neutral kaons are not eigenstates of the CP operator.
After tremendous experimental efforts, also ``direct'' CP violation, which
is caused {\it directly} at the amplitude level through the interference between
different weak amplitudes, could be established in the neutral kaon system 
in 1999 by the NA48 (CERN) and KTeV (FNAL) collaborations,\cite{eps-prime} 
thereby ruling out superweak scenarios of CP violation.\cite{superweak}
The world average taking also the final NA48 and KTeV results\cite{eps-prime-final}
into account is given as follows:
\begin{equation}\label{epsp-average}
\mbox{Re}(\varepsilon'/\varepsilon)=(16.6\pm1.6)\times10^{-4}.
\end{equation}
As far as the theoretical status of this observable is concerned, the 
short-distance contributions are under full control. On the other hand,
the long-distance part, which is described by hadronic matrix elements of 
certain four-quark operators, suffers from large uncertainties.
Although theoretical analyses performed within the Standard Model 
give results in the ball park of (\ref{epsp-average}), stringent tests
cannot be performed unless progress on the long-distance contributions
can be made.\cite{bu-ja} 

In 2001, CP-violating effects were also discovered in the $B$-meson 
system by the BaBar (SLAC) and Belle (KEK) collaborations,\cite{CP-B-obs}
representing the first observation of this phenomenon {\it outside} the 
$K$-meson system. The corresponding CP asymmetry arises in 
the ``golden'' decay $B_d\to J/\psi K_{\rm S}$,\cite{bisa} and is
induced through the interference between the 
$B^0_d\to J/\psi K_{\rm S}$ and $\bar B^0_d\to J/\psi K_{\rm S}$ decay processes 
that is caused by $B^0_d$--$\bar B^0_d$ mixing. In the summer of 2004, also 
direct CP violation could be detected by the BaBar and Belle collaborations in 
$B_d\to\pi^\mp K^\pm$ decays,\cite{CP-dir-B} thereby complementing the 
observation of this phenomenon in the neutral kaon system. 

Despite tremendous progress over the last years, we have still an incomplete 
picture of CP violation and flavour physics. The exploration of these topics is
very exciting, as it may open a window to the physics lying beyond the
Standard Model (SM), where quark-flavour physics is governed by the 
Cabibbo--Kobayashi--Maskawa (CKM) matrix.\cite{cab,KM} Indeed, in 
scenarios for new physics (NP), we typically encounter
also new sources for flavour-changing processes and CP violation. Important
examples are models with extended Higgs sectors, supersymmetric (SUSY) or
left--right-symmetric scenarios for NP. In this context, it is also important 
to note that the experimental evidence for non-vanishing neutrino masses points 
to an origin beyond the SM, raising many interesting questions, 
which include also the possibility of CP violation in the neutrino sector.\cite{altarelli}

Interestingly, CP violation plays also an outstanding r\^ole in cosmology, where this
phenomenon is one of the necessary ingredients for the generation of
the matter--antimatter asymmetry of the Universe,\cite{kolb} as was pointed
out by Sakharov in 1967.\cite{sakharov} However, model calculations show
that the CP violation present in the SM is too small to explain
this asymmetry. The required additional sources of CP violation may be associated 
with very high energy scales, as in the scenario of ``leptogenesis'', 
involving CP-violating decays of very heavy Majorana neutrinos.\cite{BPY} On the 
other hand, there are also several extensions of the SM with
new sources of CP violation that could actually be accessible in the laboratory,
as we have noted above.

Before searching for NP, we have first to understand the picture of flavour
physics emerging within the SM. Here the usual key problem for the theoretical
interpretation is related to hadronic uncertainties, where $\varepsilon'/\varepsilon$
is a famous example. In the $B$-meson system, the situation is much more
promising: it offers various strategies to explore CP violation and flavour
physics -- simply speaking, there are {\it many} $B$ decays -- and we may 
search for SM relations, which are on solid theoretical ground and 
may well be affected by NP. Concerning the kaon system,
the future lies on ``rare'' decays, which are absent at the tree level of the
SM, i.e.\ originate from loop processes, and are theoretically
very clean. A particularly important r\^ole is played by $K^+\to\pi^+\nu\bar\nu$ 
and $K_{\rm L}\to\pi^0\nu\bar\nu$, which offer poweful tests of the flavour  
sector of the SM. 

These aspects are the focus of these lectures. The outline is as follows: 
in Section~\ref{sec:CPV-SM}, we discuss the description of CP violation in 
the SM and introduce the unitarity triangle(s). We then move on to the system 
of the $B$ mesons in Section~\ref{sec:B}, where we classify non-leptonic $B$ 
decays, introduce the concept of low-energy effective Hamiltonians, and have a 
closer look at the CP-violating asymmetries arising in neutral $B$ decays. In 
Section~\ref{sec:rare}, we turn to rare decays, and discuss 
$B_{s,d}\to\mu^+\mu^-$ modes as a more detailed example. After addressing the
question of how NP may generically enter CP-violating phenomena and
rare decays in Section~\ref{sec:NP}, we are well prepared to discuss
the ``golden'' decays $B_d\to J/\psi K_{\rm S}$ and $B_s\to J/\psi \phi$ in 
Section~\ref{sec:gold}, and how we may challenge the SM through 
$B_d\to\phi K_{\rm S}$ modes in Section~\ref{sec:BphiK}. In Section~\ref{sec:puzzles},
we consider an example of a systematic strategy to search for NP, which is an
an analysis of puzzling patterns in the $B\to\pi\pi,\pi K$ data and their interplay 
with rare $K$ and $B$ decays. Finally, we conclude and give a brief outlook in 
Section~\ref{sec:concl}. 

In order to complement the discussion given here, I refer the reader to
the reviews, lecture notes and textbooks collected in 
Refs.~15--21, 
where many more details and different perspectives of the field can be found.
There are also other fascinating aspects of flavour physics and CP violation,
which are, however, beyond the scope of these lectures. Important
examples are the $D$-meson system,\cite{petrov} electric dipole 
moments,\cite{PR} or the search for flavour-violating charged 
lepton decays.\cite{CEPRT} In order to get an overview of these
topics, the reader should consult the corresponding references.

\section{CP Violation in the Standard Model}\label{sec:CPV-SM}
\subsection{Weak Interactions of Quarks}
In the SM of electroweak interactions, CP-violating effects 
are associated with the charged-current interactions of the quarks:
\begin{equation}\label{cc-int}
D\to U W^-.
\end{equation}
Here $D\in\{d,s,b\}$ and $U\in\{u,c,t\}$ denote down- and up-type quark 
flavours, respectively, whereas the $W^-$ is the usual $SU(2)_{\rm L}$ 
gauge boson. From a phenomenological point of view, it is convenient to 
collect the generic ``coupling strengths'' $V_{UD}$ of the charged-current 
processes in (\ref{cc-int}) in the form of a $3\times 3$ matrix. From a 
theoretical point of view, this ``quark-mixing'' matrix -- the CKM matrix -- 
connects the electroweak states $(d',s',b')$ of the down, strange and bottom 
quarks with their mass eigenstates $(d,s,b)$ through the following unitary 
transformation :
\begin{equation}\label{ckm}
\left(\begin{array}{c}
d'\\
s'\\
b'
\end{array}\right)=\left(\begin{array}{ccc}
V_{ud}&V_{us}&V_{ub}\\
V_{cd}&V_{cs}&V_{cb}\\
V_{td}&V_{ts}&V_{tb}
\end{array}\right)\cdot
\left(\begin{array}{c}
d\\
s\\
b
\end{array}\right)
\equiv \hat V_{\rm CKM} \cdot
\left(\begin{array}{c}
d\\
s\\
b
\end{array}\right).
\end{equation}
Consequently, $\hat V_{\rm CKM}$ is actually a {\it unitary} matrix.
This feature ensures the absence of flavour-changing neutral-current 
(FCNC) processes at the tree level in the SM, and is hence at the basis 
of the Glashow--Iliopoulos--Maiani (GIM) mechanism.\cite{GIM} 
If we express the non-leptonic charged-current interaction Lagrangian 
in terms of the mass eigenstates in (\ref{ckm}), we arrive at 
\begin{equation}\label{cc-lag2}
{L}_{\mbox{{\scriptsize int}}}^{\mbox{{\scriptsize CC}}}=
-\frac{g_2}{\sqrt{2}}\left(\begin{array}{ccc}
\bar u_{\mbox{{\scriptsize L}}},& \bar c_{\mbox{{\scriptsize L}}},
&\bar t_{\mbox{{\scriptsize L}}}\end{array}\right)\gamma^\mu\,\hat
V_{\mbox{{\scriptsize CKM}}}
\left(\begin{array}{c}
d_{\mbox{{\scriptsize L}}}\\
s_{\mbox{{\scriptsize L}}}\\
b_{\mbox{{\scriptsize L}}}
\end{array}\right)W_\mu^\dagger\,\,+\,\,\mbox{h.c.,}
\end{equation}
where $g_2$ is the $SU(2)_{\mbox{{\scriptsize L}}}$ gauge coupling, 
and the $W_\mu^{(\dagger)}$ field corresponds to the charged $W$ bosons. 
Looking at the interaction vertices following from (\ref{cc-lag2}), we observe 
that the elements of the CKM matrix describe in fact the generic
strengths of the associated charged-current processes, as we have 
noted above. 

Since the CKM matrix elements governing a $D\to U W^-$ transition and its 
CP conjugate $\bar D\to \bar U W^+$ are related to each other through
\begin{equation}\label{CKM-CP}
V_{UD}\stackrel{{ CP}}{\longrightarrow}V_{UD}^\ast,
\end{equation}
we observe that CP violation is associated with complex phases
of the CKM matrix. Consequently, the question of whether we may actually 
have {\it physical} complex phases in this matrix arises.

\subsection{Phase Structure of the CKM Matrix}
We may redefine the up- and down-type quark fields as follows:
\begin{equation}
U\to \exp(i\xi_U)U,\quad D\to \exp(i\xi_D)D. 
\end{equation}
If we perform such transformations in (\ref{cc-lag2}), the invariance 
of the charged-current interaction Lagrangian implies
\begin{equation}\label{CKM-trafo}
V_{UD}\to\exp(i\xi_U)V_{UD}\exp(-i\xi_D).
\end{equation}
Eliminating unphysical phases through these transformations, we
are left with the following parameters in the case of a general $N\times N$ 
quark-mixing matrix, where $N$ denotes the number of fermion generations:
\begin{equation}
\underbrace{\frac{1}{2}N(N-1)}_{\mbox{Euler angles}} \, + \,
\underbrace{\frac{1}{2}(N-1)(N-2)}_{\mbox{complex phases}}=
(N-1)^2.
\end{equation}

If we apply this expression to $N=2$ generations, we observe
that only one rotation angle -- the Cabibbo angle
$\theta_{\rm C}$\cite{cab} -- is required for the parametrization of the $2\times2$
quark-mixing matrix, which can be written as
\begin{equation}\label{Cmatrix}
\hat V_{\rm C}=\left(\begin{array}{cc}
\cos\theta_{\rm C}&\sin\theta_{\rm C}\\
-\sin\theta_{\rm C}&\cos\theta_{\rm C}
\end{array}\right),
\end{equation}
where $\sin\theta_{\rm C}=0.22$ follows from $K\to\pi\ell\bar\nu_\ell$ 
decays. On the other hand, in the case of $N=3$ generations, the 
parametrization of the corresponding $3\times3$ quark-mixing matrix involves 
three Euler-type angles and a single {\it complex} phase. This complex phase 
allows us to accommodate CP violation in the SM, as was pointed out by 
Kobayashi and Maskawa in 1973.\cite{KM} The corresponding picture
is referred to as the Kobayashi--Maskawa (KM) mechanism of CP violation.

In the ``standard parametrization'' advocated by the Particle Data Group,\cite{PDG} 
the three-generation CKM matrix takes the following form:
\begin{equation}\label{standard}
\hat V_{\rm CKM}=
\left(\begin{array}{ccc}
c_{12}c_{13}&s_{12}c_{13}&s_{13}e^{-i\delta_{13}}\\ -s_{12}c_{23}
-c_{12}s_{23}s_{13}e^{i\delta_{13}}&c_{12}c_{23}-
s_{12}s_{23}s_{13}e^{i\delta_{13}}&
s_{23}c_{13}\\ s_{12}s_{23}-c_{12}c_{23}s_{13}e^{i\delta_{13}}&-c_{12}s_{23}
-s_{12}c_{23}s_{13}e^{i\delta_{13}}&c_{23}c_{13}
\end{array}\right),
\end{equation}
where $c_{ij}\equiv\cos\theta_{ij}$ and $s_{ij}\equiv\sin\theta_{ij}$. 
If we redefine the quark-field phases appropriately, $\theta_{12}$, $\theta_{23}$ 
and $\theta_{13}$ can all be made to lie in the first quadrant. The advantage of 
this parametrization is that the mixing between two generations $i$ and $j$ vanishes 
if $\theta_{ij}$ is set to zero. In particular, for 
$\theta_{23}=\theta_{13}=0$, the third generation decouples, and the
submatrix describing the mixing between the first and 
second generations takes the same form as (\ref{Cmatrix}).

\subsection{Wolfenstein Parametrization}\label{ssec:wolf}
The charged-current interactions of the quarks exhibit an interesting hierarchy,
which follows from experimental data:\cite{PDG} transitions within the same 
generation involve CKM matrix elements of $O(1)$, those between the first and the second generation are associated with CKM elements of $O(10^{-1})$, those between the second and the third generation are related to CKM elements of $O(10^{-2})$, 
and those between the first and third generation are described by CKM matrix 
elements of $O(10^{-3})$. For phenomenological applications, it would be useful 
to have a parametrization of the CKM matrix available that makes this pattern 
explicit.\cite{wolf} To this end, we introduce a set of new parameters, 
$\lambda$, $A$, $\rho$ and $\eta$, by imposing the following 
relations:\cite{blo}
\begin{equation}\label{set-rel}
s_{12}\equiv\lambda=0.22,\quad s_{23}\equiv A\lambda^2,\quad 
s_{13}e^{-i\delta_{13}}\equiv A\lambda^3(\rho-i\eta).
\end{equation}
Going back to the standard parametrization (\ref{standard}), we 
obtain an {\it exact} parametrization of the CKM matrix as a function of 
$\lambda$ (and $A$, $\rho$, $\eta$), which allows us to expand each CKM 
element in powers of the small parameter $\lambda$. Neglecting terms of 
${O}(\lambda^4)$ yields the ``Wolfenstein 
parametrization'':\cite{wolf}
\begin{equation}\label{W-par}
\hat V_{\mbox{{\scriptsize CKM}}} =\left(\begin{array}{ccc}
1-\frac{1}{2}\lambda^2 & \lambda & A\lambda^3(\rho-i\eta) \\
-\lambda & 1-\frac{1}{2}\lambda^2 & A\lambda^2\\
A\lambda^3(1-\rho-i\eta) & -A\lambda^2 & 1
\end{array}\right)+{O}(\lambda^4).
\end{equation}

\subsection{Unitarity Triangle(s)}
The unitarity of the CKM matrix, which is described by
\begin{equation}
\hat V_{\mbox{{\scriptsize CKM}}}^{\,\,\dagger}\cdot\hat 
V_{\mbox{{\scriptsize CKM}}}=
\hat 1=\hat V_{\mbox{{\scriptsize CKM}}}\cdot\hat V_{\mbox{{\scriptsize 
CKM}}}^{\,\,\dagger},
\end{equation}
leads to a set of 12 equations, consisting of 6 normalization 
and 6 orthogonality relations. The latter can be represented as 6 
triangles in the complex plane, all having the same area, which
represents a measure of the ``strenghth'' of CP violation in
the SM.

\begin{figure}[t]
\centerline{
\begin{tabular}{lr}
   \epsfysize=3.3truecm
   \epsffile{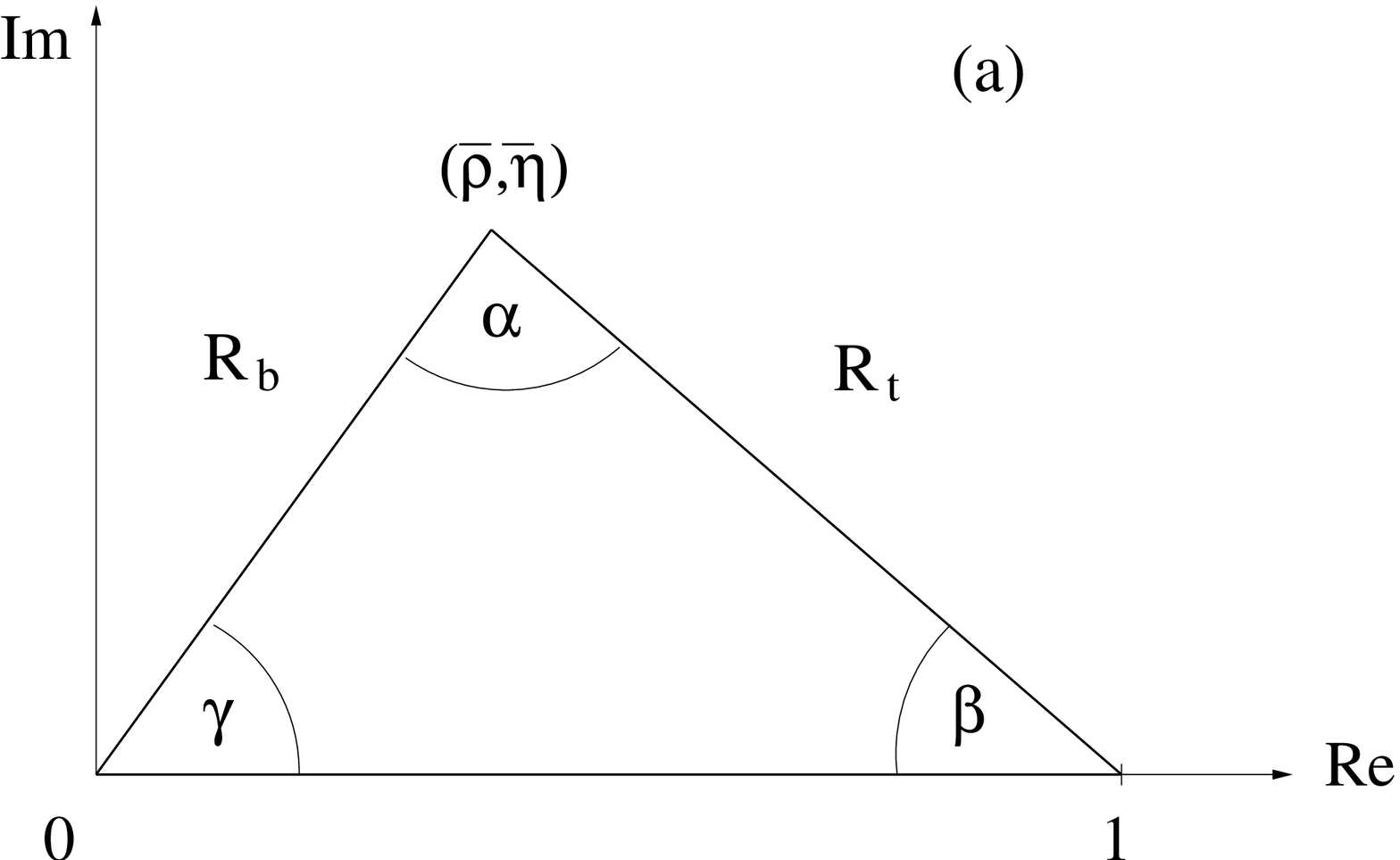}
&
   \epsfysize=3.3truecm
   \epsffile{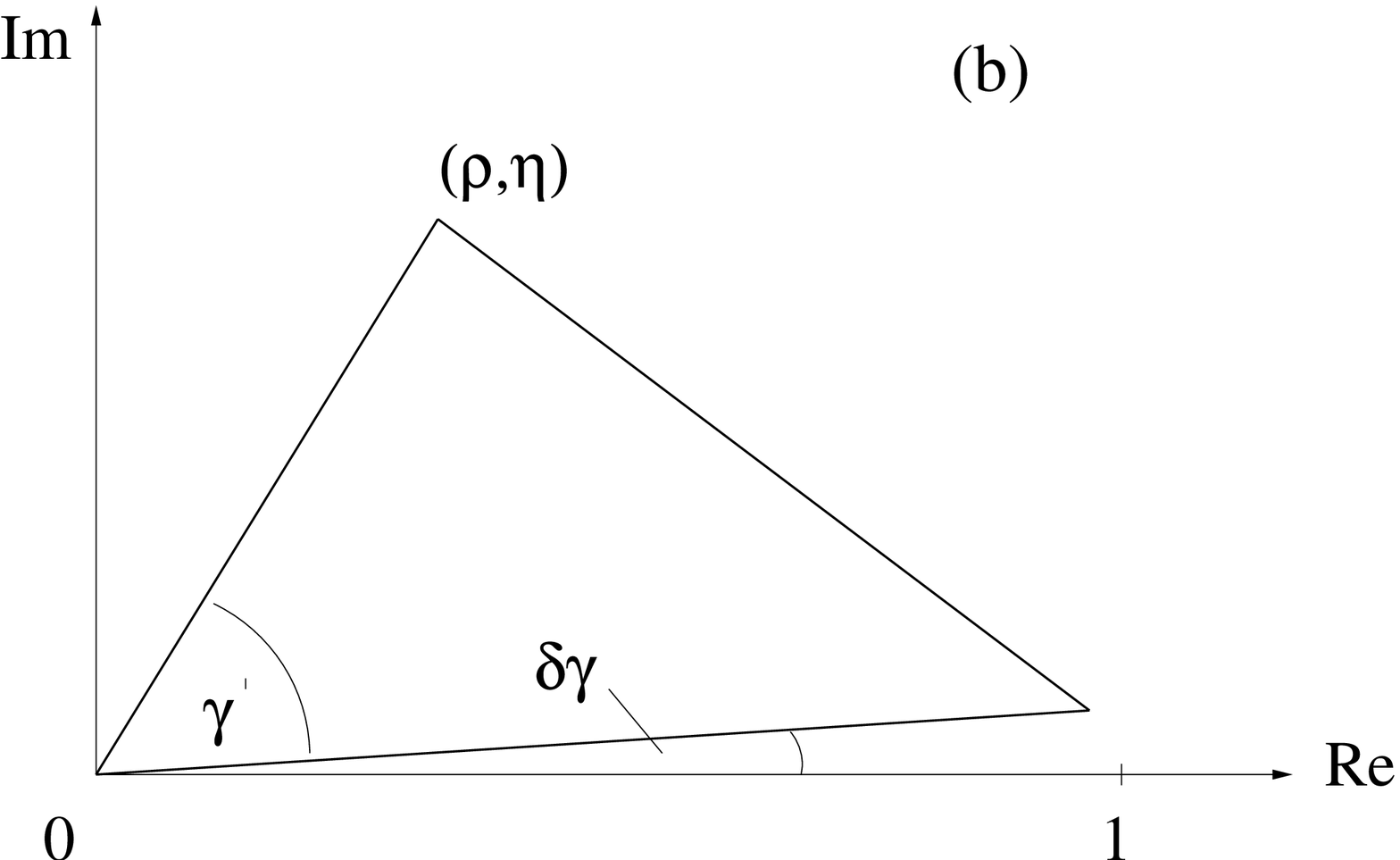}
\end{tabular}}
\caption{The two non-squashed unitarity triangles of the CKM matrix, as
explained in the text: (a) and (b) correspond to the orthogonality 
relations (\ref{UT1}) and (\ref{UT2}), respectively.}
\label{fig:UT}
\end{figure}

Using the Wolfenstein parametrization of the CKM matrix, the
generic shape of these triangles can be explored. Interestingly,
only the following two orthogonality relations correspond to the 
case of triangles, where all three sides are of the same order of
magnitude:
\begin{eqnarray}
V_{ud}V_{ub}^\ast+V_{cd}V_{cb}^\ast+V_{td}V_{tb}^\ast & = &
0\label{UT1}\\
V_{ud}^\ast V_{td}+V_{us}^\ast V_{ts}+V_{ub}^\ast V_{tb}
& = & 0;\label{UT2}
\end{eqnarray}
in the other triangles, one side is suppressed with respect to the
others by factors of $O(\lambda^2)$ or $O(\lambda^4)$. If we
apply the Wolfenstein parametrization by keeping just the leading, 
non-vanishing terms of the expansion in $\lambda$, (\ref{UT1})
and (\ref{UT2}) give the same result, which is given by
\begin{equation}
\left[(\rho+i\eta)+(1-\rho-i\eta)+(-1)\right]A\lambda^3=0,
\end{equation}
and describes {\it the} unitarity triangle of the CKM matrix. Taking also the
next-to-leading order corrections in $\lambda$ into account,\cite{blo} 
as described in Subsection~\ref{ssec:wolf}, we arrive at the triangles illustrated in
Fig.~\ref{fig:UT}. The apex of the triangle in Fig.~\ref{fig:UT} (a) is simply given by 
\begin{equation}\label{rho-eta-bar}
\bar\rho\equiv\rho\left[1-\frac{1}{2}\lambda^2\right],\quad
\bar\eta\equiv\eta\left[1-\frac{1}{2}\lambda^2\right],
\end{equation}
corresponding to the triangle sides
\begin{equation}\label{Rb-Rt-def}
R_b
\equiv\left[1-\frac{\lambda^2}{2}\right]\frac{1}{\lambda}\left|\frac{V_{ub}}{V_{cb}}\right|,
\quad
R_t
\equiv\frac{1}{\lambda}\left|\frac{V_{td}}{V_{cb}}\right|.
\end{equation}
Obviously, this triangle is the straightforward generalization of the
leading-order case, and is usually considered in the literature. Whenever referring
to {\it a} unitarity triangle (UT) in the following discussion, we shall always mean this
triangle. On the other hand, the characteristic feature of the triangle in 
Fig.~\ref{fig:UT} (b) is that $\gamma=\gamma'+\delta\gamma$, with
\begin{equation}
\delta\gamma=\lambda^2\eta=O(1^\circ).
\end{equation}

\begin{figure}[t] 
   \centering
   \includegraphics[width=10.0truecm]{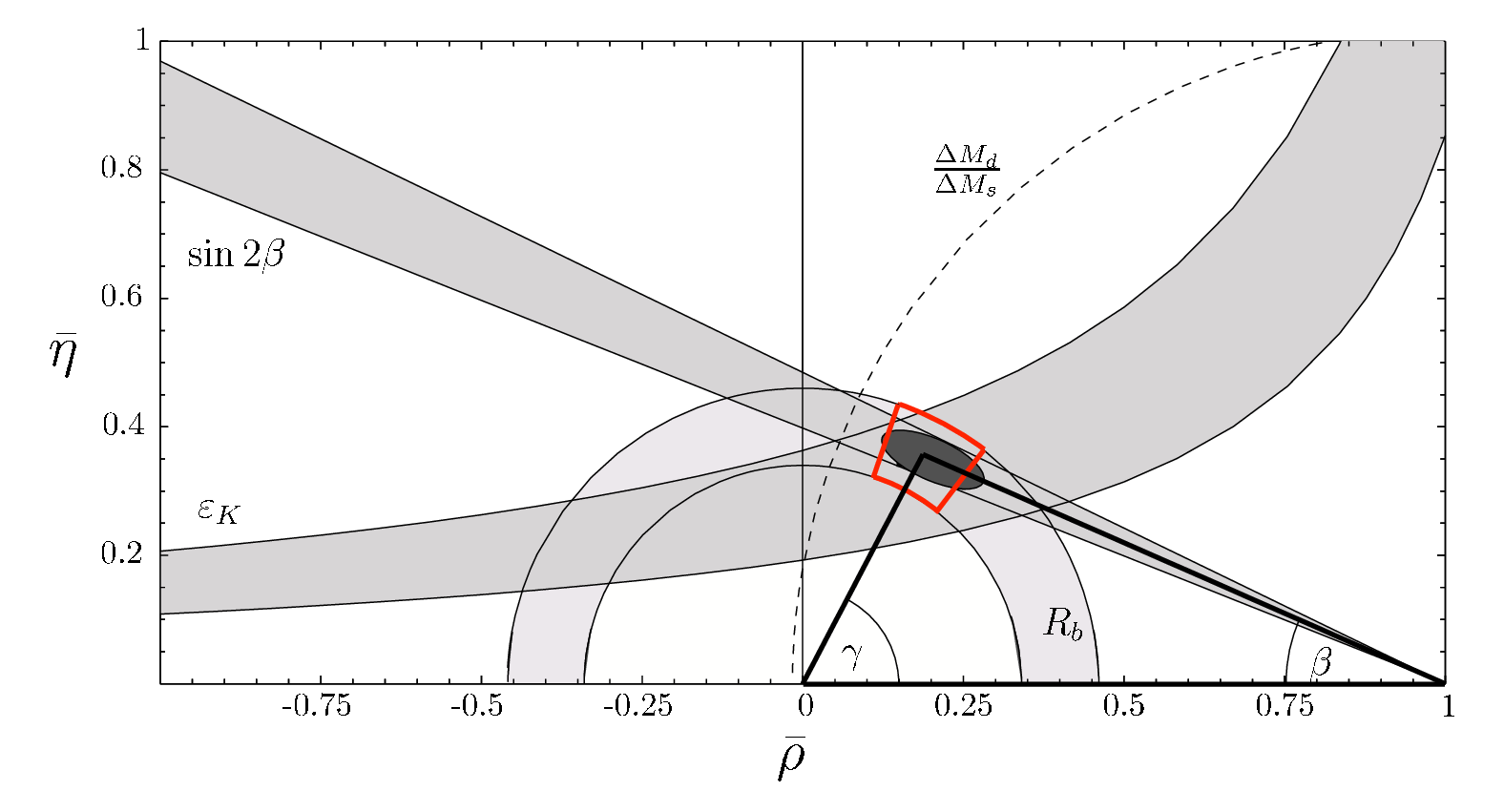} 
   \vspace*{-0.3truecm}
   \caption{The current situation in the $\bar\rho$--$\bar\eta$ 
   plane, as discussed in the text.}\label{fig:UT-fit}
\end{figure}

\subsection{Determination of the Unitarity Triangle}\label{ssec:UT}
There are two conceptually different avenues to determine the UT: 
\begin{itemize}
\item[(i)] In the ``CKM fits'', theory is used to convert 
experimental data into contours in the $\bar\rho$--$\bar\eta$ plane, where 
semileptonic $b\to u \ell \bar\nu_\ell$, $c \ell \bar\nu_\ell$ decays and 
$B^0_{d,s}$--$\bar B^0_{d,s}$ mixing (see \ref{ssec:B-basic}) 
allow us to determine the UT sides 
$R_b$ and $R_t$, respectively, i.e.\ to fix two circles in the $\bar\rho$--$\bar\eta$ 
plane. On the other hand, the indirect CP violation in the neutral kaon system
described by $\varepsilon_K$ can be transformed into a hyperbola. 
\item[(ii)] Theory allows us to convert measurements of CP-violating effects in 
$B$-meson decays into direct information on the UT angles. The most prominent
example is the determination of $\sin2\beta$ through $B_d\to J/\psi K_{\rm S}$,
but several other strategies were proposed.
\end{itemize}
The goal is to ``overconstrain'' the UT as much as possible. In the future, 
additional contours can be fixed in the $\bar\rho$--$\bar\eta$ plane through 
the measurement of rare decays. For example, BR$(K^+\to\pi^+\nu\bar\nu)$
can be converted into an ellipse, and BR$(K_{\rm L}\to\pi^0\nu\bar\nu)$
allows the determination of $|\bar\eta|$. 

In Fig.~\ref{fig:UT-fit}, we show the current situation: the shaded dark ellipse 
is the result of a CKM fit,\cite{BSU} the straight
lines represent the measurement of $\sin 2\beta$ (see Subsection~\ref{ssec:BpsiK}), 
and the quadrangle corresponds to a determination of $\gamma$ from 
$B_d\to\pi^+\pi^-$, $B_d\to\pi^\mp K^\pm$  
decays,\cite{BFRS-update} which will be discussed in Section~\ref{sec:puzzles}. 
For very comprehensive analyses of the UT, we refer the reader to the web sites of the 
``CKM Fitter Group'' and the ``UTfit collaboration''.\cite{CKM-UT-fits} 

The overall consistency with the SM is very impressive. Furthermore, 
also the recent data for $B\to\pi\rho,\rho\rho$ as well as 
$B_d\to D^{(*)\pm}\pi^\mp$ and $B\to D K$ decays give constraints for 
the UT that are in accordance with the KM mechanism, although the errors are 
still pretty large in several of these cases. Despite this remarkably consistent
picture, there is still hope to encounter deviations from the SM. Since 
$B$ mesons play a key r\^ole in this adventure, 
let us next have a closer look at them.

\section{System of the $B$ Mesons}\label{sec:B}
\subsection{Basic Features}\label{ssec:B-basic}
In this decade, there are promising perspectives for the exploration
of $B$-meson decays: the asymmetric $e^+$--$e^-$ $B$ factories at SLAC and
KEK, with their detectors BaBar and Belle, respectively, are taking 
data since several years and could already produce $O(10^8)$ $B\bar B$ 
pairs. Moreover, the CDF and D0 collaborations have recently reported the 
first results from run II of Fermilab's Tevatron. Starting in 2007, the 
LHC\cite{lef} at 
CERN will allow ``second-generation'' $B$-decay studies through the
dedicated LHCb experiment, and also ATLAS and CMS can address 
certain interesting aspects of $B$ physics. For the more distant future,
an $e^+$--$e^-$ ``super-$B$ factory'' is under consideration, with an
increase of luminosity by two orders of magnitude with respect to 
the currently operating machines.\cite{super-B}

The $B$-meson system offers also a very interesting playground
for theorists, involving exciting aspects of strong and weak interactions,
as well as the possible impact of physics beyond the SM.
Moreover, there is an extremely fruitful interplay between theory and
experiment in this field, and despite impressive progress, there are
still aspects left that could not yet be accessed experimentally and
are essentially unexplored.

\begin{figure}
   \centering
   \vspace*{-0.1truecm}
   \includegraphics[width=5.9truecm]{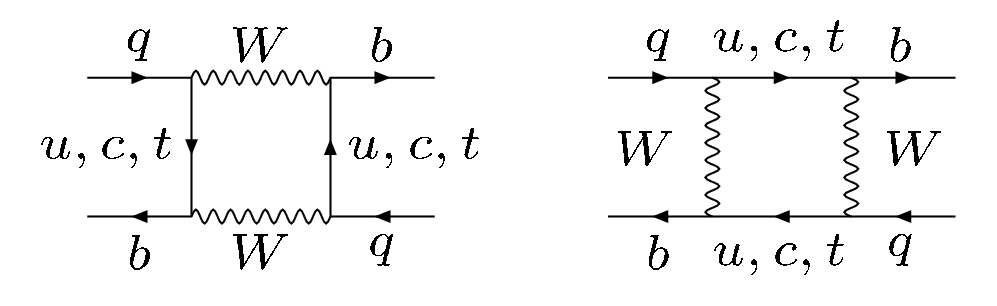} 
   \vspace*{-0.2truecm}
   \caption{Box diagrams contributing to $B^0_q$--$\bar B^0_q$ mixing in the
   SM ($q\in\{d,s\}$).}
   \label{fig:boxes}
\end{figure}

The $B$-meson system consists of charged and neutral mesons, which are 
characterized by the following valence-quark contents:
\begin{displaymath}
\mbox{charged:}~\left[\begin{array}{ccc}
B^+\sim u\,\bar b, && B^-\sim \bar u\,b\\
B^+_c\sim c\,\bar b, && B^+_c\sim\bar c\,b
\end{array}\right],
\quad
\mbox{neutral:}~\left[\begin{array}{ccc}
B^0_d\sim d\,\bar b, && \bar B^0_d\sim\bar d\,b \\
B^0_s\sim s\,\bar b, && \bar B^0_s\sim\bar s\,b
\end{array}\right].
\end{displaymath}
The characteristic feature
of the neutral $B_q$ ($q\in \{d,s\}$) mesons is 
$B_q^0$--$\bar B_q^0$ mixing, which we encountered already
in the determination of the UT discussed in
Subsection~\ref{ssec:UT}. In the SM, this phenomenon,
which is the counterpart of $K^0$--$\bar K^0$ mixing, originates from 
box diagrams, as illustrated in Fig.~\ref{fig:boxes}. Due to 
$B_q^0$--$\bar B_q^0$ mixing, an initially, i.e.\ at time $t=0$,  present 
$B^0_q$-meson state evolves into the following time-dependent linear 
combination:
\begin{equation}
|B_q(t)\rangle=a(t)|B^0_q\rangle+
b(t)|\bar B^0_q \rangle.
\end{equation}
The coefficients $a(t)$ and $b(t)$ are governed by an appropriate Schr\"odinger
equation, with mass eigenstates that are characterized by mass and decay
width differences $\Delta M_q$ and $\Delta\Gamma_q$, respectively. The
time-dependent transition rates for decays of initially present $B^0_q$ or
$\bar B^0_q$ mesons into a final state $f$ involve $\cos(\Delta M_qt)$ and
$\sin(\Delta M_qt)$ terms, describing the $B_q^0$--$\bar B_q^0$
oscillations.

\begin{figure}
   \centering
   \includegraphics[width=3.0truecm]{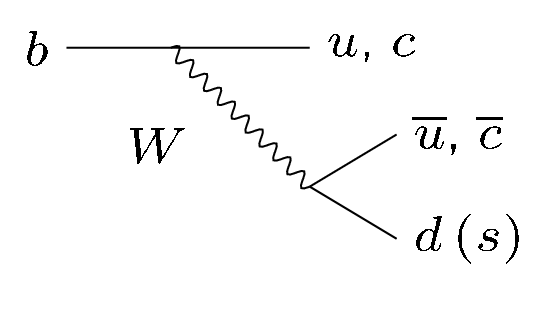} 
   \vspace*{-0.2truecm}
   \caption{Tree diagrams ($q_1,q_2\in\{u,c\}$).}
   \label{fig:tree}
\end{figure}

\begin{figure}
   \centering
   \vspace*{-0.6truecm}
   \includegraphics[width=4.7truecm]{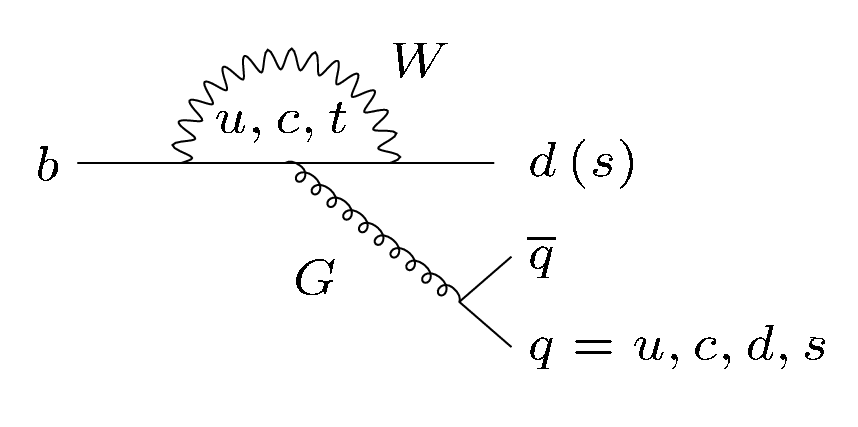} 
   \vspace*{-0.2truecm}
   \caption{QCD penguin diagrams ($q_1=q_2\in\{u,d,c,s\}$).}
   \label{fig:QCD-pen}
\end{figure}

\begin{figure}
   \centering
   \vspace*{-0.6truecm}
   \includegraphics[width=8.0truecm]{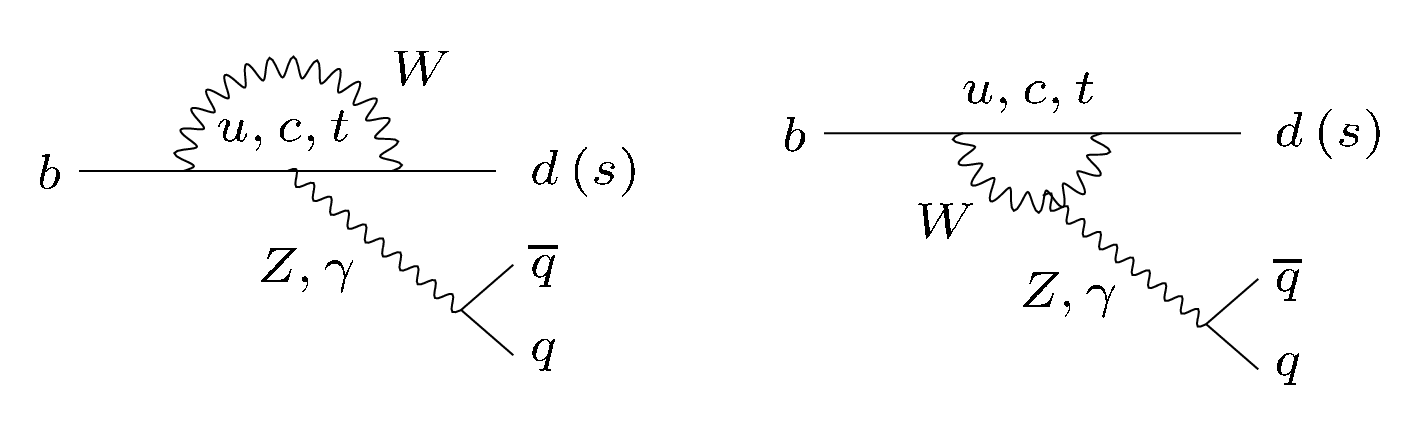} 
   \vspace*{-0.2truecm}
   \caption{Electroweak penguin diagrams 
($q_1=q_2\in\{u,d,c,s\}$).}
   \label{fig:EW-pen}
\end{figure}

\subsection{Classification of Non-Leptonic $B$ Decays}\label{ssec:non-lept-class}
For the exploration of CP violation, non-leptonic $B$ decays play the
key r\^ole. The final states of such transitions consist only of
quarks, and they are mediated by $b\to q_1\,\bar q_2\,d\,(s)$ quark-level 
processes, with $q_1,q_2\in\{u,d,c,s\}$. There are two kinds of 
topologies contributing to such decays: ``tree'' and ``penguin'' 
topologies. The latter consist of gluonic (QCD) and electroweak (EW) 
penguins. In Figs.~\ref{fig:tree}--\ref{fig:EW-pen}, the corresponding 
leading-order Feynman diagrams are shown. Depending
on the flavour content of their final states, we may classify the 
non-leptonic $b\to q_1\,\bar q_2\,d\,(s)$ decays as follows:
\begin{itemize}
\item $q_1\not=q_2\in\{u,c\}$: {\it only} tree diagrams contribute.
\item $q_1=q_2\in\{u,c\}$: tree {\it and} penguin diagrams contribute.
\item $q_1=q_2\in\{d,s\}$: {\it only} penguin diagrams contribute.
\end{itemize}

\subsection{Low-Energy Effective Hamiltonians}\label{ssec:eff-ham}
\subsubsection{General Structure}
For the analysis of non-leptonic $B$ decays, we use 
low-energy effective Hamiltonians, which are calculated by making use 
of the ``operator product expansion'', yielding transition amplitudes
of the following structure:
\begin{equation}\label{LE-Ham}
\langle f|{H}_{\rm eff}|i\rangle=\frac{G_{\rm F}}{\sqrt{2}}
\lambda_{\rm CKM}\sum_k C_k(\mu)\langle f|Q_k(\mu)|i\rangle.
\end{equation}
Here $G_{\rm F}$ denotes Fermi's constant, $\lambda_{\rm CKM}$ is a
CKM factor, and $\mu$ denotes a renormalization scale. The technique of the 
operator product expansion allows us to separate the short-distance 
contributions to this transition amplitude from the long-distance ones, 
which are described by perturbative quantities $C_k(\mu)$ (``Wilson 
coefficient functions'') and non-perturbative quantities 
$\langle f|Q_k(\mu)|i\rangle$ (``hadronic matrix elements''), respectively. 
The $Q_k$ are local operators, which are generated through the
electroweak interactions and the interplay with QCD, and govern 
``effectively'' the decay in question. The Wilson
coefficients are -- simply speaking -- the scale-dependent couplings
of the vertices described by the $Q_k$.

\subsubsection{Illustration through an Example}
Let us consider the quark-level process $b\to\ c\bar u s$, which 
originates from a tree diagram of the kind shown in Fig.~\ref{fig:tree}, 
as a simple illustration. If we ``integrate out'' the $W$ boson having
four-momentum $k$, i.e.\ use the relation
\begin{equation}
\frac{g_{\nu\mu}}{k^2-M_W^2}
\quad\stackrel{k^2 \ll M_W^2}{\longrightarrow}\quad
-\,\frac{g_{\nu\mu}}{M_W^2}\equiv-\left(\frac{8G_{\rm F}}{\sqrt{2}g_2^2}
\right)g_{\nu\mu},
\end{equation}
we arrive at the following low-energy effective Hamiltonian:
\begin{equation}
{H}_{\rm eff}=\frac{G_{\rm F}}{\sqrt{2}}V_{us}^\ast V_{cb} O_2,
\end{equation}
with the ``current--current" operator
\begin{equation}
O_2\equiv\left[\bar s_\alpha\gamma_\mu(1-\gamma_5)u_\alpha\right]
\left[\bar c_\beta\gamma^\mu(1-\gamma_5)b_\beta\right]
\end{equation}
and the Wilson coefficient $C_2=1$; $\alpha$ and $\beta$ are the 
$SU(3)_{\rm C}$ indices of QCD. Taking now QCD effects, i.e.\ the
exchange of gluons, into account and performing a proper ``matching'' 
between the full and the effective theories, a second current--current operator,
\begin{equation}
O_1\equiv\left[\bar s_\alpha\gamma_\mu(1-\gamma_5)u_\beta\right]
\left[\bar c_\beta\gamma^\mu(1-\gamma_5)b_\alpha\right],
\end{equation}
is generated, involving a Wilson coefficient $C_1(\mu)$. Due to 
the impact of QCD, also the Wilson coefficient of $O_2$ acquires 
now a renormalization-scale dependence and deviates from one. 
The results for the $C_k(\mu)$ contain
terms of $\log(\mu/M_W)$, which become large for $\mu=O(m_b)$, the typical
scale governing the hadronic matrix elements of the four-quark operators
$O_k$. In order to deal with these large logarithms, ``renormalization-group-improved''
perturbation theory offers the appropriate tool. The fact that 
$\langle f|H_{\rm eff}|i\rangle$ in (\ref{LE-Ham}) {\it cannot}
depend on the renormalization scale $\mu$ implies a renormalization group
equation, which has a solution of the following form:
\begin{equation}\label{RG-evol}
\vec C(\mu)=\hat U (\mu, M_W)\cdot \vec C(M_W).
\end{equation}
Here the ``evolution matrix'' $\hat U (\mu, M_W)$ connects the initial 
values $\vec C(M_W)$ encoding the whole {\it short-distance} physics at
high-energy scales with the coefficients at scales at the level of a few
GeV. Following these lines, 
\begin{equation}
\alpha_s^n\left[\log\left(\frac{\mu}{M_W}\right)\right]^n \mbox{(LO)}, \quad 
\alpha_s^n\left[\log\left(\frac{\mu}{M_W}\right)\right]^{n-1} \mbox{(NLO)}, \quad ...
\end{equation}
can be systematically summed up, where ``LO'' and ``NLO'' stand for the leading
and next-to-leading order approximations, respectively. 
For more detailed discussions, we refer the reader to 
Refs.~17, 34.     

\subsubsection{A Closer Look at Non-Leptonic Decays}
Low-energy effective Hamiltonians provide a general tool for the theoretical
description of weak $B$- and $K$-meson decays, as well as $B^0_q$--$\bar B^0_q$
and $K^0$--$\bar K^0$ mixing. Let us discuss the application to non-leptonic
$B$ decays in more detail. For the exploration of CP violation, transitions with
$\Delta C=\Delta U=0$ are particularly interesting. As can be seen from 
Figs.~\ref{fig:tree}--\ref{fig:EW-pen}, these decays receive contributions
both from tree and from penguin topologies. If we apply the unitarity of the 
CKM matrix, we find that the corresponding CKM factors are related through
\begin{equation}\label{UT-rel}
V_{ur}^\ast V_{ub}+V_{cr}^\ast V_{cb}+
V_{tr}^\ast V_{tb}=0,
\end{equation}
where $r\in\{d,s\}$. Consequently, only {\it two} independent weak amplitudes 
contribute to any given decay of this kind. In comparison with our previous 
example, which was a pure tree decay, we have now also to deal with penguin
topologies, involving -- in addition to the $W$ boson -- the top quark as a 
second ``heavy'' particle. Once these degrees of freedom are ``integrated out'', 
their influence is only felt through the initial 
conditions of the renormalization group evolution (\ref{RG-evol}).
Mathematically, the penguin topologies in Figs.~\ref{fig:QCD-pen}
and \ref{fig:EW-pen} with internal top-quark exchanges (as well
as the corresponding box diagrams in Fig.~\ref{fig:boxes}) that enter these 
coefficients are described by certain ``Inami--Lim functions''.\cite{IL} 
Finally, using (\ref{UT-rel}) to eliminate $V_{tr}^\ast V_{tb}$, 
we obtain an effective Hamiltonian of the following structure:
\begin{equation}\label{e4}
{H}_{\mbox{{\scriptsize eff}}}=\frac{G_{\mbox{{\scriptsize 
F}}}}{\sqrt{2}}\Biggl[\sum\limits_{j=u,c}V_{jr}^\ast V_{jb}\biggl\{
\sum\limits_{k=1}^2C_k(\mu)\,Q_k^{jr}+\sum\limits_{k=3}^{10}
C_k(\mu)\,Q_k^{r}\biggr\}\Biggr].
\end{equation}
Here we have introduced another quark-flavour label $j\in\{u,c\}$,
and the four-quark operators $Q_k^{jr}$ can be divided as follows:
\begin{itemize}
\item Current--current operators:
\begin{equation}
\begin{array}{rcl}
Q_{1}^{jr}&=&(\bar r_{\alpha}j_{\beta})_{\mbox{{\scriptsize V--A}}}
(\bar j_{\beta}b_{\alpha})_{\mbox{{\scriptsize V--A}}}\\
Q_{2}^{jr}&=&(\bar r_\alpha j_\alpha)_{\mbox{{\scriptsize 
V--A}}}(\bar j_\beta b_\beta)_{\mbox{{\scriptsize V--A}}}.
\end{array}
\end{equation}
\item QCD penguin operators:
\begin{equation}\label{qcd-penguins}
\begin{array}{rcl}
Q_{3}^r&=&(\bar r_\alpha b_\alpha)_{\mbox{{\scriptsize V--A}}}\sum_{q'}
(\bar q'_\beta q'_\beta)_{\mbox{{\scriptsize V--A}}}\\
Q_{4}^r&=&(\bar r_{\alpha}b_{\beta})_{\mbox{{\scriptsize V--A}}}
\sum_{q'}(\bar q'_{\beta}q'_{\alpha})_{\mbox{{\scriptsize V--A}}}\\
Q_{5}^r&=&(\bar r_\alpha b_\alpha)_{\mbox{{\scriptsize V--A}}}\sum_{q'}
(\bar q'_\beta q'_\beta)_{\mbox{{\scriptsize V+A}}}\\
Q_{6}^r&=&(\bar r_{\alpha}b_{\beta})_{\mbox{{\scriptsize V--A}}}
\sum_{q'}(\bar q'_{\beta}q'_{\alpha})_{\mbox{{\scriptsize V+A}}}.
\end{array}
\end{equation}
\item EW penguin operators (the $e_{q'}$ denote the
electrical quark charges):
\begin{equation}
\begin{array}{rcl}
Q_{7}^r&=&\frac{3}{2}(\bar r_\alpha b_\alpha)_{\mbox{{\scriptsize V--A}}}
\sum_{q'}e_{q'}(\bar q'_\beta q'_\beta)_{\mbox{{\scriptsize V+A}}}\\
Q_{8}^r&=&
\frac{3}{2}(\bar r_{\alpha}b_{\beta})_{\mbox{{\scriptsize V--A}}}
\sum_{q'}e_{q'}(\bar q_{\beta}'q'_{\alpha})_{\mbox{{\scriptsize V+A}}}\\
Q_{9}^r&=&\frac{3}{2}(\bar r_\alpha b_\alpha)_{\mbox{{\scriptsize V--A}}}
\sum_{q'}e_{q'}(\bar q'_\beta q'_\beta)_{\mbox{{\scriptsize V--A}}}\\
Q_{10}^r&=&
\frac{3}{2}(\bar r_{\alpha}b_{\beta})_{\mbox{{\scriptsize V--A}}}
\sum_{q'}e_{q'}(\bar q'_{\beta}q'_{\alpha})_{\mbox{{\scriptsize V--A}}}.
\end{array}
\end{equation}
\end{itemize}
At a renormalization scale $\mu={O}(m_b)$, the Wilson coefficients of the 
current--current operators are $C_1(\mu)={O}(10^{-1})$ and $C_2(\mu)={O}(1)$, 
whereas those of the penguin operators are as large as ${O}(10^{-2})$.\cite{BBL-B}

The short-distance part of (\ref{e4}) is nowadays under full control. On the other 
hand, the long-distance piece suffers still from large theoretical uncertainties.
For a given non-leptonic decay $\bar B \to \bar f$, it is given by the hadronic 
matrix elements $\langle \bar f|Q_k(\mu)|\bar B\rangle$ of the
four-quark operators. A popular way of dealing with these quantities is to 
assume that they ``factorize'' into the product of the matrix elements of
two quark currents at some ``factorization scale'' $\mu=\mu_{\rm F}$. 
This procedure can be justified in the large-$N_{\rm C}$ approximation,\cite{largeN}
where $N_{\rm C}$ is the number of $SU(N_{\rm C})$ quark colours, 
and there are decays, where this concept can be justified because of 
``colour transparency" arguments.\cite{QCDF-old} However, it is in general 
not on solid ground. Interesting theoretical progress could be made 
 through the development of the 
``QCD factorization" (QCDF)\cite{BBNS} and ``perturbative QCD" 
(PQCD)\cite{PQCD} approaches, the soft collinear effective theory 
(SCET),\cite{SCET} and QCD light-cone sum-rule methods.\cite{LCSR} 
An important target of these methods is given by $B\to\pi\pi$ and
$B\to\pi K$ decays. Thanks to the $B$ factories, the corresponding 
theoretical results can now be confronted with experiment. Since the
data indicate large non-factorizable 
corrections,\cite{BFRS-update,BFRS}$^{\mbox{--}}$\cite{CGRS} 
the long-distance contributions to these decays remain a theoretical challenge.

\subsection{Towards the Exploration of CP Violation}
\subsubsection{Direct CP Violation}\label{ssec:CP-dir}
Let us now have a closer look at the amplitude structure of non-leptonic
$B$ decays. Because of the unitarity of the CKM matrix, at most two weak 
amplitudes contribute to such modes in the SM.
Consequently, the corresponding transition amplitudes can be written
as follows:
\begin{eqnarray}
A(\bar B\to\bar f)&=&e^{+i\varphi_1}
|A_1|e^{i\delta_1}+e^{+i\varphi_2}|A_2|e^{i\delta_2}\label{par-ampl}\\
A(B\to f)&=&e^{-i\varphi_1}|A_1|e^{i\delta_1}+
e^{-i\varphi_2}|A_2|e^{i\delta_2}.\label{par-ampl-CP}
\end{eqnarray}
Here the $\varphi_{1,2}$ denote CP-violating weak phases, originating
from the CKM matrix, whereas the 
$|A_{1,2}|e^{i\delta_{1,2}}$ are CP-conserving ``strong'' amplitudes, 
which contain the whole hadron dynamics of the decay at hand:
\begin{equation}\label{ampl-struc}
|A_j|e^{i\delta_j}\sim\sum\limits_k
C_{k}(\mu) \langle\bar f|Q_{k}^j(\mu)|\bar B\rangle.
\end{equation}
Using (\ref{par-ampl}) and (\ref{par-ampl-CP}), we obtain
the following CP asymmetry:
\begin{eqnarray}
{A}_{\rm CP}&\equiv&\frac{\Gamma(B\to f)-
\Gamma(\bar B\to\bar f)}{\Gamma(B\to f)+\Gamma(\bar B
\to \bar f)}=\frac{|A(B\to f)|^2-|A(\bar B\to \bar f)|^2}{|A(B\to f)|^2+
|A(\bar B\to \bar f)|^2}\nonumber\\
&=&\frac{2|A_1||A_2|\sin(\delta_1-\delta_2)
\sin(\varphi_1-\varphi_2)}{|A_1|^2+2|A_1||A_2|\cos(\delta_1-\delta_2)
\cos(\varphi_1-\varphi_2)+|A_2|^2}.\label{direct-CPV}
\end{eqnarray}
We observe that a non-vanishing value can be generated through the 
interference between the two weak amplitudes, provided both a non-trivial 
weak phase difference $\varphi_1-\varphi_2$ and a non-trivial strong phase 
difference $\delta_1-\delta_2$ are present. This kind of
CP violation is referred to as ``direct'' CP violation, as it originates 
directly at the amplitude level of the considered decay. It is the 
$B$-meson counterpart of the effect that is probed through 
$\mbox{Re}(\varepsilon'/\varepsilon)$ in the neutral kaon 
system,\footnote{For the calculation of $\mbox{Re}(\varepsilon'/\varepsilon)$, 
an approriate low-energy effective Hamiltonian with the same structure as (\ref{e4}) 
is used. The large theoretical uncertainties mentioned after (\ref{epsp-average}) 
originate from a strong cancellation between the QCD and EW penguin
contributions (caused by the large top-quark mass), and the associated
hadronic matrix elements.} and could recently be established 
with the help of $B_d\to\pi^\mp K^\pm$ 
decays,\cite{CP-dir-B} as we will see in Subsection~\ref{ssec:tiny-EWP}.

\subsubsection{Strategies}
Since $\varphi_1-\varphi_2$ is in general given by one of the angles of 
the UT -- usually $\gamma$ -- the goal is to extract this
quantity from the measured value of ${A}_{\rm CP}$. Unfortunately, 
hadronic uncertainties enter this game through the poorly known 
hadronic matrix elements in (\ref{ampl-struc}). In order to
deal with this problem, we may proceed along one of the 
following two avenues:
\begin{itemize}
\item[(i)] Amplitude relations can be used to eliminate the 
hadronic matrix elements. We distinguish between exact relations, 
using pure ``tree'' decays  of the kind $B\to KD$ or $B_c\to D_sD$, 
and relations, which follow from the flavour symmetries of strong interactions, 
i.e.\ isospin or $SU(3)_{\rm F}$, and involve $B_{(s)}\to\pi\pi,\pi K,KK$ modes. 
\item[(ii)] In decays of neutral $B_q$ mesons ($q\in\{d,s\}$), interference effects 
between $B^0_q$--$\bar B^0_q$ mixing and decay processes may induce
 ``mixing-induced CP violation''. If a single CKM amplitude governs the decay, 
 the hadronic matrix elements cancel in the corresponding
CP asymmeties; otherwise we have to use again amplitude relations.
\end{itemize}

\subsubsection{CP Violation in Neutral $B_q$ Decays}\label{ssec:B-neut}
Since neutral  $B_q$ mesons are a key element for the exploration of
CP violation, let us next have a closer look at their most important features.
A particularly simple -- but also very interesting  -- situation arises in
decays into final states $f$ that are eigenstates of the CP operator, i.e.\ satisfy
\begin{equation}
\hat C\hat P|f\rangle=\pm |f\rangle. 
\end{equation}
If we solve the Schr\"odinger equation describing $B^0_q$--$\bar B^0_q$
mixing as we noted in Subsection~\ref{ssec:B-basic}, we obtain the following 
time-dependent CP asymmetry:
\begin{eqnarray}
\lefteqn{\left.\frac{\Gamma(B^0_q(t)\to f)-
\Gamma(\bar B^0_q(t)\to f)}{\Gamma(B^0_q(t)\to f)+
\Gamma(\bar B^0_q(t)\to f)}\right|_{\Delta\Gamma_q=0}}\nonumber\\
&&={A}_{\rm CP}^{\rm dir}(B_q\to f)\,\cos(\Delta M_q t)+
{A}_{\rm CP}^{\rm mix}(B_q\to f)\,\sin(\Delta M_q t).
\end{eqnarray}
Here the coefficient of the $\cos(\Delta M_q t)$ term is given by
\begin{equation}
{A}^{\mbox{{\scriptsize dir}}}_{\mbox{{\scriptsize CP}}}(B_q\to f)=
\frac{|A(B^0_q\to f)|^2-|A(\bar B^0_q\to \bar f)|^2}{|A(B^0_q\to f)|^2+
|A(\bar B^0_q\to \bar f)|^2},
\end{equation}
and measures the direct CP violation in the decay
$B_q\to f$. As we have seen in (\ref{direct-CPV}), this phenomenon 
originates from the interference between different weak amplitudes. 
On the other hand, the coefficient of the $\sin(\Delta M_q t)$ term 
describes another kind of CP violation, which is caused by the interference 
between $B^0_q$--$\bar B^0_q$ mixing and decay processes, and
is referred to as ``mixing-induced" CP violation. Mathematically, it is
described by
\begin{equation}\label{CPV-mix}
{A}^{\mbox{{\scriptsize mix}}}_{\mbox{{\scriptsize
CP}}}(B_q\to f)\equiv\frac{2\,\mbox{Im}\,\xi^{(q)}_f}{1+
\bigl|\xi^{(q)}_f\bigr|^2},
\end{equation}
where
\begin{equation}\label{xi-expr}
\xi_f^{(q)}=\pm e^{-i\Theta_{\rm M}^{(q)}}
\left[\frac{A(\bar B^0_q\to\bar f)}{A(B^0_q\to f)}\right]
\end{equation}
involves the CP-violating weak phase $\Theta_{\rm M}^{(q)}$ that is associated with
$B^0_q$--$\bar B^0_q$ mixing. In the SM, it is related to the CKM
phase of the box diagrams with internal top-quark exchanges shown in
Fig.~\ref{fig:boxes} as follows:
\begin{equation}
\Theta_{\rm M}^{(q)}-\pi = 2\,\mbox{arg} (V_{tq}^\ast V_{tb}) \equiv
\phi_q =  \left\{
\begin{array}{cl}
+2\beta=O(47^\circ)&~~\mbox{($q=d$)}\\
-2\delta\gamma=O(-1^\circ) &~~\mbox{($q=s$),}
\end{array}\right.
\end{equation}
where $\beta$ and $\delta\gamma$ were introduced in 
Figs.~\ref{fig:UT} (a) and (b), respectively.

If we use (\ref{par-ampl}) and (\ref{par-ampl-CP}), we may rewrite (\ref{xi-expr}) 
as follows:
\begin{equation}\label{xi-re}
\xi_f^{(q)}=\mp\, e^{-i\phi_q}\left[
\frac{e^{+i\varphi_1}|A_1|e^{i\delta_1}+
e^{+i\varphi_2}|A_2|e^{i\delta_2}}{
e^{-i\varphi_1}|A_1|e^{i\delta_1}+
e^{-i\varphi_2}|A_2|e^{i\delta_2}}\right],
\end{equation}
and observe -- in analogy to the discussion of direct CP violation 
in \ref{ssec:CP-dir} -- that this quantity suffers, in
general, also from large hadronic uncertainties. However, if one 
CKM amplitude plays the dominant r\^ole, we arrive at
\begin{equation}\label{xi-si}
\xi_f^{(q)}=\mp\, e^{-i\phi_q}\left[
\frac{e^{+i\phi_f/2}|M_f|e^{i\delta_f}}{e^{-i\phi_f/2}|M_f|e^{i\delta_f}}
\right]=\mp\, e^{-i(\phi_q-\phi_f)}.
\end{equation}
Consequently, the hadronic matrix element $|M_f|e^{i\delta_f}$ cancels
in this special case. Since the requirements for direct CP violation are obviously 
no longer satisfied, the observable 
${A}^{\mbox{{\scriptsize dir}}}_{\mbox{{\scriptsize CP}}}
(B_q\to f)$ vanishes. On the other hand, we may still have mixing-induced
CP violation. In particular, 
\begin{equation}\label{Amix-simple}
{A}^{\rm mix}_{\rm CP}(B_q\to f)=\pm\sin\phi
\end{equation}
is now governed by the CP-violating weak phase difference 
$\phi\equiv\phi_q-\phi_f$ and is {\it not} affected by hadronic 
uncertainties. The corresponding time-dependent CP asymmetry takes
then the simple form
\begin{equation}\label{Amix-t-simple}
\left.\frac{\Gamma(B^0_q(t)\to f)-
\Gamma(\bar B^0_q(t)\to \bar f)}{\Gamma(B^0_q(t)\to f)+
\Gamma(\bar B^0_q(t)\to \bar f)}\right|_{\Delta\Gamma_q=0}
=\pm\sin\phi\,\sin(\Delta M_q t),
\end{equation}
and allows an elegant determination of $\sin\phi$. In Sections~\ref{sec:gold}
and \ref{sec:BphiK}, we will see that this formalism has powerful 
applications for the search of NP.

\section{Rare Decays}\label{sec:rare}
\subsection{General Features}
The exploration of flavour physics through CP violation can nicely be 
complemented through ``rare" decays. In the SM, these processes do
{\it not} arise at the tree level, but can originate through {\it loop} effects.
Consequently, rare $B$ decays are mediated by FCNC processes of the 
kind $\bar b\to \bar s$ or $\bar b\to \bar d$, whereas rare $K$ decays 
originate from their $\bar s\to \bar d$ counterparts. Prominent examples 
of rare $B$ decays are the following exclusive channels:
\begin{itemize}
\item $B\to K^\ast\gamma$, $B\to \rho\gamma$, $...$ 
\item $B\to K^\ast\mu^+\mu^-$, $B\to \rho\mu^+\mu^-$, $...$ 
\item $B_{s,d}\to \mu^+\mu^-$. 
\end{itemize}
While the $B_{s,d}\to \mu^+\mu^-$ transitions are very clean, the former 
two decay classes suffer from theoretical uncertainties that are related to
hadronic form factors and long-distance contributions. On the other hand, the 
hadronic uncertainties are much smaller in the corresponding inclusive 
decays, $B\to X_{s,d}\gamma$ and $B\to X_{s,d}\mu^+\mu^-$, which are 
therefore more promising from the theoretical point of view, but are 
unfortunately more difficult to measure; the cleanest rare $B$ decays 
are given by $B\to X_{s,d} \nu\bar \nu$ processes. A tremendous amount of 
work went into the calculation of the branching ratio of the prominent 
$B\to X_s\gamma$ decay,\cite{BuMi} and the agreement of the experimental 
value with the SM expectation implies important constraints for the 
allowed parameter space of popular NP scenarios. The phenomenology of the kaon system includes also interesting rare decays:\cite{BF-rev,BSU}
\begin{itemize}
\item $K_{\rm L}\to\pi^0 e^+ e^-$, $K_{\rm L}\to\pi^0 \mu^+ \mu^-$
\item $K_{\rm L}\to\pi^0\nu\bar\nu$, $K^+\to\pi^+\nu\bar\nu$.
\end{itemize}

\subsection{Theoretical Description}
For the theoretical description of rare decays, low-energy 
effective Hamiltonians are used, in analogy to the analysis of non-leptonic 
$B$ decays. The structure of the corresponding transition amplitudes is therefore
similar to the one of (\ref{LE-Ham}), i.e.\ the short-distance physics is 
described by perturbatively calculable Wilson coefficient functions, 
whereas the long-distance dynamics is encoded in non-perturbative
hadronic matrix elements of local operators. It is useful to rewrite the rare-decay implementation of (\ref{LE-Ham}) as follows:\cite{PBE,BH92}
\begin{equation}\label{mmaster}
{A({\rm decay})}= 
P_0({\rm decay}) + \sum_r P_r({\rm decay} ) F_r(x_t\equiv m_t^2/M_W^2).
\end{equation}
Here $\mu=\mu_0={O}(M_W)$ was chosen, and the Wilson coefficients 
$C_k(\mu_0)$ were expressed in terms of  ``master functions'' 
$F_r(x_t)$. These quantities follow from the evaluation 
of penguin and box diagrams with heavy particles running in the loops, 
i.e.\ top and $W$ in the SM, and are related to the Inami--Lim 
functions.\cite{IL} On the other hand, the term $P_0$ summarizes the 
contributions from light internal quarks, such as the charm and up quarks. 
It should be noted that $P_0$ and $P_r$ are {\it process-dependent} quantities, 
i.e.\  depend on the hadronic matrix elements of the operators $Q_k$ for
a given decay, whereas the $F_r(x_t)$ are {\it process-independent} functions.
In Section~\ref{sec:NP}, we will return to this formalism in the context of NP. 

Rare decays have many interesting features, as discussed in several 
reviews and the references therein.\cite{BF-rev,BBL-B,BuMi,rare} Let us here 
choose $B_{s,d}\to\mu^+\mu^-$ modes as a representative example,
since these decays allow a compact presentation, belong to the 
cleanest representatives of the field of rare decays, and are an important 
element of the $B$-physics programme at the LHC.\cite{LHC-Book}

\boldmath\subsection{Example: $B_{s,d}\to\mu^+\mu^-$}
\begin{figure}[t]
   \centering
   \vspace*{-0.1truecm} 
   \includegraphics[width=5.2truecm]{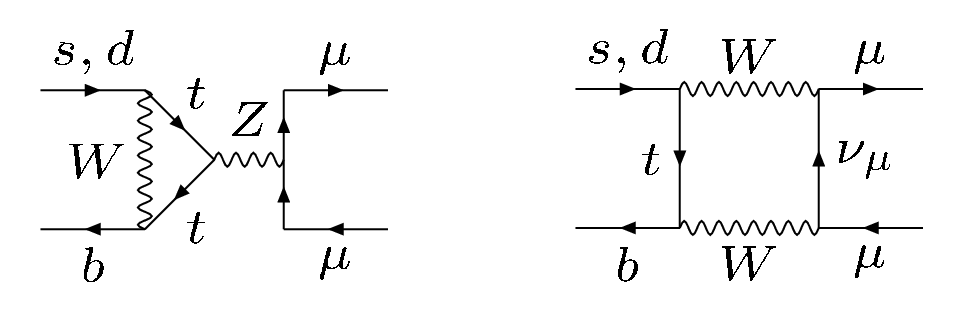}
   \vspace*{-0.1truecm} 
      \caption{Decay processes contributing to $B_{s,d}\to\mu^+\mu^-$
       in the SM.}\label{fig:Bsd-mumu-diag}
\end{figure}

\unboldmath
In the SM,  $B_q\to\mu^+\mu^-$ modes  ($q\in\{s,d\}$) originate from 
$Z^0$ penguins and box diagrams, as can be seen in Fig.~\ref{fig:Bsd-mumu-diag}.  
The corresponding low-energy effective Hamiltonian is given as follows:
\begin{equation}\label{Heff-Bmumu}
{H}_{\rm eff}=-\frac{G_{\rm F}}{\sqrt{2}}\left[
\frac{\alpha}{2\pi\sin^2\Theta_{\rm W}}\right]
V_{tb}^\ast V_{tq} \eta_Y Y_0(x_t)(\bar b q)_{\rm V-A}(\bar\mu\mu)_{\rm V-A} 
\,+\, {\rm h.c.},
\end{equation}
where $\alpha$ denotes the QED coupling, $\Theta_{\rm W}$ is the
Weinberg angle, and the short-distance physics is described by 
\begin{equation}
Y(x_t)\equiv\eta_Y Y_0(x_t).
\end{equation}
Here $\eta_Y=1.012$ is a perturbative QCD 
correction,\cite{BB-Bmumu}$\mbox{}^{\mbox{--}}$\cite{MiU} 
and the Inami--Lim function $Y_0(x_t)$, which can be written to a good 
approximation as
\begin{equation}
Y_0(x_t)=0.98\times\left[\frac{m_t}{167 \, \mbox{GeV}}\right]^{1.56},
\end{equation}
describes the top-quark mass dependence.\cite{Buras-Cracow}
We observe that the matrix element of (\ref{Heff-Bmumu}) between a 
$\langle\mu^-\mu^+|$ final state and a $|B^0_q\rangle$ initial state 
involves simply the ``decay constant'' $f_{B_q}$, which is defined 
through\footnote{Note that 
$\langle 0|\bar b \gamma_\alpha q|B^0_q(k)\rangle=0$, since the $B^0_q$ 
is a pseudoscalar meson.} 
\begin{equation}
\langle 0|\bar b \gamma_\alpha\gamma_5 q|B^0_q(k)\rangle=
i f_{B_q}k_\alpha.
\end{equation}
Consequently, we encounter a very favourable situation with respect to
the hadronic matrix elements. Since, moreover, NLO QCD corrections were 
calculated, and long-distance contributions are expected to play a negligible 
r\^ole,\cite{BB-Bmumu} the $B_q\to\mu^+\mu^-$ modes belong to the cleanest 
rare $B$ decays.

In the SM, their branching ratios can be written
as\cite{Brev01}
\begin{eqnarray}
\lefteqn{\mbox{BR}( B_s \to \mu^+ \mu^-) = 4.1 \times 10^{-9}}\nonumber\\
&&\qquad\qquad\times \left[\frac{f_{B_s}}{0.24 \, \mbox{GeV}} \right]^2 \left[
\frac{|V_{ts}|}{0.040} \right]^2 \left[
\frac{\tau_{B_s}}{1.5 \, \mbox{ps}} \right] \left[ \frac{m_t}{167 
\, \mbox{GeV} } \right]^{3.12}\label{BR-Bsmumu}\\
\lefteqn{\mbox{BR}(B_d \to \mu^+ \mu^-) = 1.1 \times 10^{-10}}\nonumber\\
&&\qquad\qquad\times\left[ \frac{f_{B_d}}{0.20 \, \mbox{GeV}} \right]^2 \left[
\frac{|V_{td}|}{0.008} \right]^2
\left[ \frac{\tau_{B_d}}{1.5 \, \mbox{ps}} \right] \left[
\frac{m_t}{167 \, \mbox{GeV} } \right]^{3.12},\label{BR-Bdmumu}
\end{eqnarray}
which should be compared with the current experimental upper bounds:
\begin{eqnarray}
\mbox{BR}( B_s \to \mu^+ \mu^-)& < & 5.0 \times 10^{-7}~\mbox{[D0
@ 95\% C.L.\cite{D0-Bsmumu}]}\\
\mbox{BR}( B_d \to \mu^+ \mu^-) & < & 8.3 \times 10^{-8}~\mbox{[BaBar 
@ 90\% C.L.\cite{BaBar-Bdmumu}].}
\end{eqnarray}
If we use the relation
\begin{equation}
R_t\equiv\frac{1}{\lambda}\left|\frac{V_{td}}{V_{cb}}\right|=
\frac{1}{\lambda}\left|\frac{V_{td}}{V_{ts}}\right|\left[1+O(\lambda^2)\right],
\end{equation}
we observe that the measurement of the ratio
\begin{equation}\label{RT1-rare}
\frac{\mbox{BR}(B_d\to\mu^+\mu^-)}{\mbox{BR}(B_s\to\mu^+\mu^-)}=
\left[\frac{\tau_{B_d}}{\tau_{B_s}}\right]
\left[\frac{M_{B_d}}{M_{B_s}}\right]
\left[\frac{f_{B_d}}{f_{B_s}}\right]^2
\left|\frac{V_{td}}{V_{ts}}\right|^2
\end{equation}
would allow a determination of the UT side $R_t$. This
strategy is complementary to the one addressed in Subsection~\ref{ssec:UT}, 
which is offered by
\begin{equation}\label{RT2-DM}
\frac{\Delta M_d}{\Delta M_s}=
\left[\frac{M_{B_d}}{M_{B_s}}\right]
\left[\frac{\hat B_{B_d}}{\hat B_{B_s}}\right]
\left[\frac{f_{B_d}}{f_{B_s}}\right]^2
\left|\frac{V_{td}}{V_{ts}}\right|^2,
\end{equation}
where the $\hat B_{B_q}$ are non-perturbative ``bag" parameters arising in
$B^0_q$--$\bar B^0_q$ mixing. These determinations rely
on the following $SU(3)$-breaking ratios:
\begin{equation}
\frac{f_{B_s}}{f_{B_d}},  \quad 
\xi\equiv\frac{\sqrt{\hat B_s}f_{B_s}}{\sqrt{\hat B_d}f_{B_d}},
\end{equation}
which can be obtained from QCD lattice studies or with the help of
QCD sum rules, and are an important target of current research.\cite{CKM-book}
Looking at (\ref{RT1-rare}) and (\ref{RT2-DM}), we see that these expressions 
imply the relation
\begin{equation}\label{Bmumu-DM-rel}
\frac{\mbox{BR}(B_s\to\mu^+\mu^-)}{\mbox{BR}(B_d\to\mu^+\mu^-)}=
\left[\frac{\tau_{B_s}}{\tau_{B_d}}\right]
\left[\frac{\hat B_{B_d}}{\hat B_{B_s}}\right]
\left[\frac{\Delta M_s}{\Delta M_d}\right],
\end{equation}
which suffers from theoretical uncertainties that are smaller than
those affecting (\ref{RT1-rare}) and (\ref{RT2-DM}) since the 
dependence on $(f_{B_d}/f_{B_s})^2$ cancels and 
$\hat B_{B_d}/\hat B_{B_s}=1$ up to tiny $SU(3)$-breaking 
corrections.\cite{Buras-rel}
Moreover, we may also use the (future) experimental data for 
$\Delta M_{(s)d}$ to reduce the hadronic uncertainties of the SM 
predictions of the $B_q\to\mu^+\mu^-$ branching 
ratios:
\begin{eqnarray}
\mbox{BR}(B_s \to \mu^+ \mu^-) &=& (3.42 \pm 0.53)\times
\left[\frac{\Delta M_s}{18.0\, {\rm ps}^{-1}}\right]\times 10^{-9}\\
\mbox{BR}(B_d\to \mu^+ \mu^-) &=& (1.00 \pm 0.14)\times 10^{-10}.
\end{eqnarray}

In view of these tiny branching ratios, we could only hope 
to observe the $B_q \to \mu^+ \mu^-$ decays at the LHC, should they 
actually be governed by their SM contributions.\cite{LHC-Book}
However, as these transitions are mediated by rare FCNC processes,
they are sensitive probes of NP. In particular, as was recently
reviewed,\cite{buras-NP} the $B_q \to \mu^+ \mu^-$ branching 
ratios may be dramatically enhanced in specific NP (SUSY) scenarios.
Should this actually be the case, these decays may be seen at run II 
of the Tevatron, and the $e^+e^-$ $B$ factories could observe 
$B_d\to \mu^+ \mu^-$. 

The interpretation of the present and future experimental constraints 
on $B_s\to\mu^+\mu^-$ in the context of the constrained minimal extension 
of the SM (CMSSM) with universal 
scalar masses was recently critically  discussed.\cite{EOS}

\section{How Could New Physics Enter?}\label{sec:NP}
\subsection{Twofold Impact of New Physics}
In order to address the question of how NP affects flavour physics, 
we use once agian the language of the
low-energy effective Hamiltonians introduced above. There are 
then two possibilities for NP to manifest itself:\cite{buras-NP} 
\begin{itemize}
\item[(i)] NP may modify the ``strength" of the operators
arising in the SM. In this case, we obtain new
short-distance functions that depend on the NP parameters, 
such as masses of charginos, squarks, charged Higgs particles
and $\tan\bar\beta\equiv v_2/v_1$ in the MSSM. The NP 
particles enter in new box and penguin diagrams and are ``integrated out''
as the $W$ and top, so that we arrive at initial conditions for the
renormalization-group evolution (\ref{RG-evol}) of the following structure:
\begin{equation}\label{WC-NP}
C_k(\mu=M_W)\to C_k^{\rm SM} + C_k^{\rm NP},
\end{equation}
where the NP pieces $C_k^{\rm NP}$ may also involve new 
CP-violating phases that are {\it not} related to the CKM matrix.
\item[(ii)] NP may lead to an enlarged operator basis:
\begin{equation}
\{Q_k\} \to \{Q_k^{\rm SM}, Q_l^{\rm NP}\},
\end{equation}
i.e.\ operators that are absent (or strongly suppressed) in the 
SM may actually play an important r\^ole, thereby 
yielding, in general, also new sources for flavour and CP violation.
\end{itemize}

\subsection{Classification of New Physics}\label{ssec:NP-class}
After these general considerations, NP can be divided into the
following classes, as was done by Buras:\cite{buras-NP}

\vspace*{0.3truecm}

\noindent
{\it Class A:} this class describes models with ``minimal flavour violation'' (MFV), which represent the simplest extension of the SM. Here the
flavour-changing processes are still governed by the CKM matrix -- in particular
there are no new sources for CP violation --  and the only relevant operators are 
those present in the SM. If we use $v$ as an abbreviation for the
set of parameters involved, the $F_r(x_t)$ introduced in (\ref{mmaster}) 
are simply replaced by generalized functions $F_r(v)$, which involve 
only seven ``master functions", $S(v)$, $X(v)$, $Y(v)$, $Z(v)$, $E(v)$,
$D'(v)$, $E'(v)$.
In (\ref{Heff-Bmumu}), we encountered already one of them, the function $Y$, which  
characterizes rare $K$, $B$ decays with $\ell^+\ell^-$ in the final states. 
Concerning $B_q \to \mu^+ \mu^-$ decays, the NP effects can hence
be included through the simple replacement $Y(x_t) \to Y(v)$.
A similar procedure applies to the expressions for $\Delta M_q$, where a
function $S(v)$ is involved. Since the {\it same} functions enter in the
$B_s$- and $B_d$-meson cases, relations (\ref{RT1-rare}), (\ref{RT2-DM}) and
(\ref{Bmumu-DM-rel}) hold not only in the SM, but also in the whole
class of MFV models, thereby providing an interesting test of this NP scenario. 
Examples are the THDM-II and constrained MSSM if 
$\tan\bar\beta$ is not too large, as well as models with one extra universal 
dimension.

\vspace*{0.3truecm}

\noindent
{\it Class B:} in contrast to class A, new operators arise, but still no new 
CP-violating phases are present. Examples of new Dirac structures are 
$({\rm V- A})\otimes ({\rm V+A})$, $({\rm S-P})\otimes ({\rm S\pm P})$, 
$\sigma_{\mu\nu}({\rm S-P})\otimes \sigma^{\mu\nu}({\rm S - P})$, which
become relevant for $B^0_q$--$\bar B^0_q$ mixing in the MSSM with 
large $\tan\bar\beta$.

\vspace*{0.3truecm}

\noindent
{\it Class C:} in contrast to class A, the Wilson coefficients of the usual
SM operators may acquire new CP-violating phases, i.e.\
the $C_k^{\rm NP}$ in (\ref{WC-NP}) may become complex, whereas 
new operators give still negligible contributions. An example is the MSSM 
with a value of $\tan\bar\beta$ that is not too large and with non-diagonal 
elements in the squark mass matrices.

\vspace*{0.3truecm}

\noindent
{\it Class D:} this class describes the general case of physics beyond the
SM with new operators {\it and} new CP-violating phases, and is 
therefore very involved. Examples are multi-Higgs models with complex 
phases in the Higgs sector, general SUSY scenarios, models with spontaneous 
CP violation and left--right-symmetric models. 

\vspace*{0.3truecm}

\noindent
{\it Class E:} in contrast to the classes introduced above, the three-generation 
CKM matrix is now {\it not unitary}, so that the UT does not close.
An example is given by models with four generations.

\begin{figure}[t]
   \centering
   \includegraphics[width=9.0truecm]{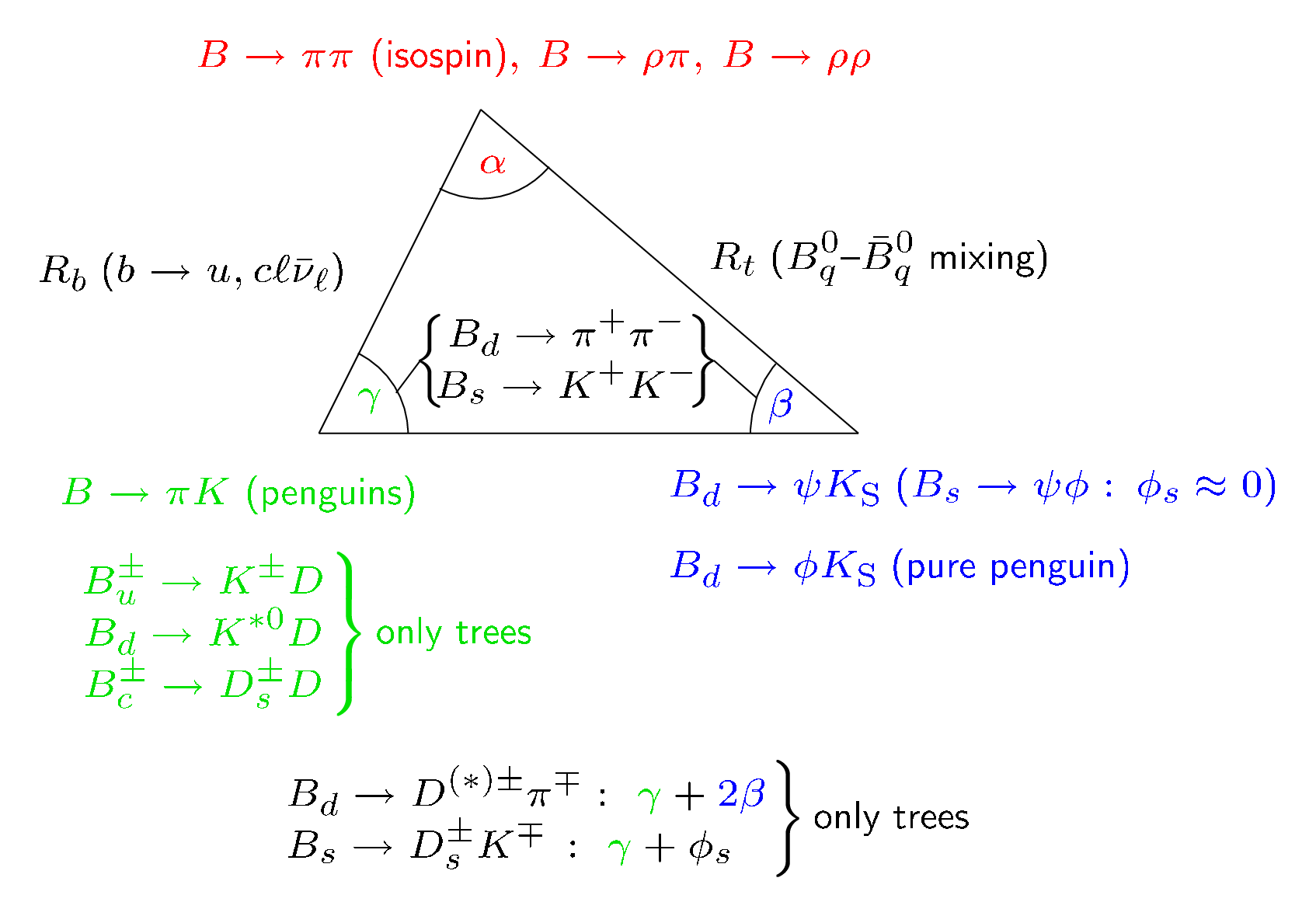} 
   \caption{A brief roadmap of $B$-decay strategies for the exploration of
   CP violation.}
   \label{fig:flavour-map}
\end{figure}

\subsection{Impact on the Roadmap of Quark-Flavour Physics}\label{ssec:road-map}
The $B$-meson system offers a variety of processes and strategies for the
exploration of CP violation.\cite{RF-Phys-Rep} 
Looking at Fig.~\ref{fig:flavour-map}, where we 
have collected prominent examples, we see that there are processes with 
a very {\it different} dynamics that are -- within the framework of the SM --
sensitive to the {\it same} angles of the UT. Moreover, rare $B$- and 
$K$-meson decays, which originate from loop effects in the SM, provide 
complementary insights into flavour physics and interesting correlations with 
the CP-B sector. 

In the presence of NP, the subtle interplay between different processes is 
expected to be disturbed, so that discrepancies should emerge. There are 
two popular avenues for NP to enter the roadmap of flavour physics:
\begin{itemize}
\item[(i)]{\it $B^0_q$--$\bar B^0_q$ mixing:} NP may enter through the exchange
of new particles in the box diagrams, or through new contributions at the
tree level, thereby modifying the mixing parameters as follows:
\begin{equation}\label{Dm-Phi-NP}
\Delta M_q=\Delta M_q^{\rm SM}+\Delta M_q^{\rm NP}, \quad
\phi_q=\phi_q^{\rm SM}+\phi_q^{\rm NP}.
\end{equation}
Whereas $\Delta M_q^{\rm NP}$ would affect the determination of the UT 
side $R_t$, the NP contribution $\phi_q^{\rm NP}$ would enter the mixing-induced
CP asymmetries. Using dimensional arguments borrowed from effective field theory,\cite{FM-BpsiK,FIM} it can be shown that 
$\Delta M_q^{\rm NP}/\Delta M_q^{\rm SM}\sim1$ and
$\phi_q^{\rm NP}/\phi_q^{\rm SM}\sim1$ may -- in principle -- be possible
for a NP scale $\Lambda_{\rm NP}$ in the TeV regime; such a pattern may 
also arise in specific NP scenarios. However,
thanks to the $B$-factory data, the space for NP is getting
smaller and smaller in the $B_d$-meson system. On the other hand, the
$B_s$ sector is still essentially unexplored, and leaves a lot of hope for the
LHC era. In Section~\ref{sec:gold}, we will discuss the corresponding
``golden'' decays, $B_d\to J/\psi K_{\rm S}$ and $B_s\to J/\psi \phi$. 
\vspace*{0.3truecm}
\item[(ii)]{\it Decay amplitudes:} NP has typically a small effect if SM tree processes
play the dominant r\^ole, as in $B_d\to J/\psi K_{\rm S}$ decays.
On the other hand, there are potentially large effects in the FCNC sector.
For instance, new particles may enter in penguin diagrams, or we may encounter 
new FCNC contributions at the tree level. Sizeable contributions may arise
generically in field-theoretical estimates with 
$\Lambda_{\rm NP}\sim\mbox{TeV}$,\cite{FM-BphiK} as well as in
specific NP models.  Interestingly, there are hints in the current
$B$-factory data that this may actually be the case. In particular, Belle results for 
the $B_d\to \phi K_{\rm S}$ channel raise the question of whether 
$(\sin2\beta)_{\phi K_{\rm S}}=(\sin2\beta)_{\psi K_{\rm S}}$, 
and the branching ratios of certain $B\to\pi K$ decays show a puzzling pattern
which may indicate NP in the EW penguin sector. These hot topics will be 
discussed in Sections~\ref{sec:BphiK} and \ref{sec:puzzles}, respectively.
\end{itemize}
\vspace*{0.1truecm}
Let us emphasize that also the $D$-meson system provides 
interesting probes for the search of NP:\cite{petrov}  $D^0$--$\bar D^0$ mixing 
and CP-violating effects are tiny in the SM, but may be enhanced 
through NP. 

Concerning model-dependent NP analyses, in particular SUSY scenarios have
received a lot of attention; for a selection of recent studies, see
Refs.~65--70.   
Examples of other fashionable NP scenarios are left--right-symmetric 
models,\cite{LR-sym} scenarios with extra dimensions,\cite{extra-dim} 
models with an extra $Z'$,\cite{Z-prime} little Higgs scenarios,\cite{little-higgs}  
and models with a fourth generation.\cite{hou-4}

\section{``Golden'' Decays of $B$ Mesons}\label{sec:gold}
Let us now have a closer look at $B_d\to J/\psi K_{\rm S}$ and 
$B_s\to J/\psi \phi$, which are important 
applications of the formalism discussed in Subsection~\ref{ssec:B-neut}.
\subsection{$B_d\to J/\psi K_{\rm S}$}\label{ssec:BpsiK}
\subsubsection{Amplitude Structure and CP-Violating Observables}
This decay has a CP-odd final state, and originates from 
$\bar b\to\bar c c \bar s$ quark-level transitions. Consequently,
as we have seen in the classification of Subsection~\ref{ssec:non-lept-class},
we have to deal both with tree and with penguin topologies, so that
the decay amplitude takes the following form:\cite{RF-BdsPsiK}
\begin{equation}\label{Bd-ampl1}
A(B_d^0\to J/\psi K_{\rm S})=\lambda_c^{(s)}\left(A_{\rm T}^{c'}+
A_{\rm P}^{c'}\right)+\lambda_u^{(s)}A_{\rm P}^{u'}
+\lambda_t^{(s)}A_{\rm P}^{t'}.
\end{equation}
In this expression, the
\begin{equation}\label{lamqs-def}
\lambda_q^{(s)}\equiv V_{qs}V_{qb}^\ast
\end{equation}
are CKM factors, $A_{\rm T}^{c'}$ is the CP-conserving strong tree amplitude, 
while the $A_{\rm P}^{q'}$ describe the penguin topologies with internal 
$q$-quark exchanges ($q\in\{u,c,t\})$, including QCD and EW penguins; 
the primes remind us that we are dealing with a $\bar b\to\bar s$ 
transition. If we eliminate now $\lambda_t^{(s)}$ with the help of (\ref{UT-rel}) 
and apply the Wolfenstein parametrization, we arrive at 
\begin{equation}\label{BdpsiK-ampl2}
A(B_d^0\to J/\psi K_{\rm S})\propto\left[1+\lambda^2 a e^{i\theta}
e^{i\gamma}\right],
\end{equation}
where
\begin{equation}
a e^{i\vartheta}\equiv\left(\frac{R_b}{1-\lambda^2}\right)
\left[\frac{A_{\rm P}^{u'}-A_{\rm P}^{t'}}{A_{\rm T}^{c'}+
A_{\rm P}^{c'}-A_{\rm P}^{t'}}\right]
\end{equation}
is a hadronic parameter. 

Using now the formalism of Subsection~\ref{ssec:B-neut} we obtain
\begin{equation}\label{xi-BdpsiKS}
\xi_{\psi K_{\rm S}}^{(d)}=+e^{-i\phi_d}\left[\frac{1+
\lambda^2a e^{i\vartheta}e^{-i\gamma}}{1+\lambda^2a e^{i\vartheta}
e^{+i\gamma}}\right].
\end{equation}
Unfortunately, $a e^{i\vartheta}$, which is a measure for the ratio of the
$B_d^0\to J/\psi K_{\rm S}$ penguin to tree contributions,
can only be estimated with large hadronic uncertainties. However, since 
this parameter enters (\ref{xi-BdpsiKS}) in a doubly Cabibbo-suppressed way, its 
impact on the CP-violating observables is practically negligible. We can put 
this important statement on a more quantitative basis by making the plausible
assumption that $a={O}(\bar\lambda)={O}(0.2)={O}(\lambda)$,
where $\bar\lambda$ is a ``generic'' expansion parameter:
\begin{eqnarray}
{A}^{\mbox{{\scriptsize dir}}}_{\mbox{{\scriptsize
CP}}}(B_d\to J/\psi K_{\mbox{{\scriptsize S}}})&=&0+
{O}(\overline{\lambda}^3)\label{Adir-BdpsiKS}\\
{A}^{\mbox{{\scriptsize mix}}}_{\mbox{{\scriptsize
CP}}}(B_d\to J/\psi K_{\mbox{{\scriptsize S}}})&=&-\sin\phi_d +
{O}(\overline{\lambda}^3) \, \stackrel{\rm SM}{=} \, -\sin2\beta+
{O}(\overline{\lambda}^3).\label{Amix-BdpsiKS}
\end{eqnarray}
Consequently, (\ref{Amix-BdpsiKS}) allows an essentially {\it clean}
determination of $\sin2\beta$.\cite{bisa}

\subsubsection{Experimental Status}
Since the CKM fits performed within the SM pointed to a large value of 
$\sin2\beta$, $B_d\to J/\psi K_{\rm S}$ offered the exciting perspective 
of {\it large} mixing-induced CP violation. In 2001, the measurement of  
${A}^{\mbox{{\scriptsize mix}}}_{\mbox{{\scriptsize
CP}}}(B_d\to J/\psi K_{\mbox{{\scriptsize S}}})$
allowed indeed the first observation of CP violation {\it outside} the 
$K$-meson system.\cite{CP-B-obs}
The most recent data are still not showing any signal for {\it direct} CP violation
in $B_d\to J/\psi K_{\rm S}$ decays, as is expected from (\ref{Adir-BdpsiKS}),
but yield
\begin{equation}
\sin2\beta=\left\{
\begin{array}{ll}
0.722\pm0.040\pm0.023 & \mbox{(BaBar\cite{s2b-babar})}\\
0.728\pm0.056\pm0.023 & \mbox{(Belle\cite{s2b-belle}),}
\end{array}
\right.
\end{equation}
which gives the following world average:\cite{HFAG}
\begin{equation}\label{s2b-average}
\sin 2\beta=0.725\pm0.037.
\end{equation}
The theoretical uncertainties are below the 0.01 level (a recent analysis finds even
smaller effects\cite{BMR}), and can be controlled
in the LHC era with the help of the $B_s\to J/\psi K_{\rm S}$ channel.\cite{RF-BdsPsiK}

\subsubsection{What about New Physics?}
The agreement of (\ref{s2b-average}) with the CKM fits is 
excellent.\cite{CKM-UT-fits} However, despite this remarkable 
feature, NP could -- in principle -- still be hiding in the mixing-induced 
CP violation observed in $B_d\to J/\psi K_{\rm S}$. The point is that the
key quantity is actually the $B^0_d$--$\bar B^0_d$ mixing phase
\begin{equation}
\phi_d=\phi_d^{\rm SM}+\phi_d^{\rm NP}=2\beta+\phi_d^{\rm NP},
\end{equation}
where the world average (\ref{s2b-average}) implies
\begin{equation}\label{phid-exp}
\phi_d=(46.5^{+3.2}_{-3.0})^\circ \quad\lor\quad 
(133.5^{+3.0}_{-3.2})^\circ.
\end{equation}
Here the former solution would be in excellent agreement with the CKM fits, yielding
$40^\circ\lsim2\beta\lsim50^\circ$, whereas the latter would correspond to 
NP.\cite{FIM,FlMa} Both solutions can be distinguished through the measurement 
of the sign of $\cos\phi_d$, where a positive value would select the SM case. 
Performing an angular analysis of 
$B_d\to J/\psi[\to\ell^+\ell^-] K^\ast[\to\pi^0K_{\rm S}]$ processes,
the BaBar collaboration finds\cite{babar-c2b}
\begin{equation}
\cos\phi_d =2.72^{+0.50}_{-0.79} \pm 0.27,
\end{equation}
thereby favouring the SM. Interestingly, this picture emerges also 
from the first data for CP-violating effects in $B_d\to D^{(*)\pm}\pi^\mp$ 
modes,\cite{RF-BDpi} and an analysis of $B\to\pi\pi,\pi K$ decays,\cite{BFRS} 
although in an indirect manner.

As far as NP contributions at the amplitude level are concerned, they have 
to compete with SM tree-diagram-like topologies, which play the dominant
r\^ole in the $B\to J/\psi K $ modes. Consequently, the NP contributions to
the decay amplitudes are generically at most at the 10\% level; these effects
could be detected through appropriate observables, exploiting direct
CP violation and charged $B^\pm\to J/\psi K^\pm$ decays.\cite{FM-BpsiK}
Since the current $B$-factory data do not give any indication for NP of this kind,
we eventually arrive at the situation in the $\bar\rho$--$\bar\eta$ plane 
shown in Fig.~\ref{fig:UT-fit}. The space for NP contributions to 
$B^0_d$--$\bar B^0_d$ mixing is therefore getting smaller and smaller. 
However, there is still hope for NP effects in $B^0_s$--$\bar B^0_s$ mixing, 
which can nicely be probed through $B_s\to J/\psi \phi$, the $B_s$-meson
counterpart of $B_d\to J/\psi K_{\rm S}$.

\subsection{$B_s\to J/\psi \phi$}\label{ssec:Bspsiphi}
\subsubsection{Preliminaries: Characteristic Features of $B^0_s$--$\bar B^0_s$
Mixing}
At the $e^+e^-$ $B$ factories operating at the $\Upsilon(4S)$ resonance, 
{\it no} $B_s$ mesons are accessible, whereas we obtain plenty of $B_s$ 
mesons at hadron colliders, i.e.\ at Tevatron-II and the LHC. The
$B_s$ system has interesting features:
\begin{itemize}
\item In the SM, the $B^0_s$--$\bar B^0_s$ oscillations are expected to be much faster
than their $B_d$-meson counterparts, and could so far not be observed.
The current lower bound for the mass difference of the $B_s$ mass
eigenstates is given as follows:\cite{HFAG}
\begin{equation}
\left.\Delta M_s\right|_{\rm SM}>14.4 \, \mbox{ps}^{-1}~\mbox{(95\% C.L.),}
\end{equation}
and plays an important r\^ole in the CKM fits.\cite{CKM-book}

\item In contrast to the $B_d$-meson system, the width difference $\Delta\Gamma_s$
is expected to be sizable in the $B_s$ case,\cite{lenz} 
and may therefore allow interesting
studies with the following ``untagged'' $B_s$ rates:\cite{dun,FD}
\begin{equation}
\langle\Gamma(B_q(t)\to f)\rangle\equiv \Gamma(B_q^0(t)\to f)+
\Gamma(\bar B_q^0(t)\to f).
\end{equation}
Recently, the first results for $\Delta\Gamma_s$ were reported from the 
Tevatron, using the $B_s\to J/\psi\phi$ channel:\cite{DDF}
\begin{equation}
\frac{|\Delta\Gamma_s|}{\Gamma_s}=\left\{
\begin{array}{ll}
0.65^{+0.25}_{-0.33}\pm0.01 & \mbox{(CDF\cite{CDF-DG})}\\
0.21^{+0.33}_{-0.45} & \mbox{(D0\cite{D0-DG})}.
\end{array}
\right.
\end{equation}

\item Finally, let us emphasize again that
$\phi_s$ is negligibly small in the SM, whereas $\phi_d$ takes the 
large value of $2\beta=(46.5^{+3.2}_{-3.0})^\circ$.
\end{itemize}

\subsubsection{CP Violation in $B_s\to J/\psi\phi$}
This channel is simply related to $B_d\to J/\psi K_{\rm S}$ through a
replacement of the down spectator quark by a strange quark. Consequently,
the structure of the $B_s\to J/\psi\phi$ decay amplitude is  
completely analogous to (\ref{BdpsiK-ampl2}). On the other hand, the 
final state of  $B_s\to J/\psi\phi$ is an admixture of different CP eigenstates, 
which can, however, be disentangled through an angular 
analysis.\cite{DDF,DDLR} The corresponding angular distribution 
exhibits tiny direct CP violation, and allows the extraction of
\begin{equation}\label{sinphis}
\sin\phi_s+{O}(\overline{\lambda}^3)=\sin\phi_s+{O}(10^{-3})
\end{equation}
through mixing-induced CP violation.
Since $\phi_s=-2\lambda^2\eta={O}(10^{-2})$ in the SM, the 
determination of this phase from (\ref{sinphis}) is affected by
hadronic uncertainties of ${O}(10\%)$, which may become an
important issue for the LHC era. These uncertainties can be controlled with
the help of flavour-symmetry arguments through the decay 
$B_d\to J/\psi \rho^0$.\cite{RF-ang} 

Because of its nice experimental signature, $B_s\to J/\psi\phi$ is very accessible
at hadron colliders, and can be fully exploited at the LHC.
Needless to note, the big hope is that {\it sizeable} CP violation
will be found in this channel. Since the CP-violating effects in 
$B_s\to J/\psi\phi$ are tiny in the SM, this would give us an unambiguous
signal for NP.\cite{NiSi} As the situation for NP entering 
through the decay amplitude is similar to $B\to J/\psi K$, we would get 
immediate evidence for NP contributions to $B^0_s$--$\bar B^0_s$ mixing, 
and could extract the corresponding {\it sizeable} value of $\phi_s$.\cite{DFN}
Such a scenario may generically arise in the presence of NP with 
$\Lambda_{\rm NP}\sim\mbox{TeV}$,\cite{RF-Phys-Rep} as well as 
in specific models (see, e.g., Refs.~66, 68).  
In such studies, also correlations with CP-violating effects in $B_d\to \phi K_{\rm S}$
are typically investigated, which is our next topic.

\section{Challenging the Standard Model through 
$B_d\to \phi K_{\rm S}$}\label{sec:BphiK}
\subsection{Amplitude Structure and CP-Violating Observables}
Another important probe for the testing of the KM mechanism is 
offered by $B_d^0\to \phi K_{\rm S}$, which is a 
decay into a CP-odd final state, and originates from 
$\bar b\to \bar s s \bar s$ transitions. Consequently, it is a pure penguin 
mode, which is dominated by QCD penguins.\cite{BphiK-old} Because of
the large top-quark mass, EW penguins have a sizeable impact as 
well.\cite{RF-EWP,DH-PhiK} In the SM, we may write
\begin{equation}
A(B_d^0\to \phi K_{\rm S})=\lambda_u^{(s)}\tilde A_{\rm P}^{u'}
+\lambda_c^{(s)}\tilde A_{\rm P}^{c'}+\lambda_t^{(s)}\tilde A_{\rm P}^{t'},
\end{equation}
where we have applied the same notation as in Subsection~\ref{ssec:BpsiK}.
Eliminating once more the CKM factor $\lambda_t^{(s)}$ with the help of
(\ref{UT-rel}) yields
\begin{equation}
A(B_d^0\to \phi K_{\rm S})\propto
\left[1+\lambda^2 b e^{i\Theta}e^{i\gamma}\right],
\end{equation}
where 
\begin{equation}
b e^{i\Theta}\equiv\left(\frac{R_b}{1-\lambda^2}\right)\left[
\frac{\tilde A_{\rm P}^{u'}-\tilde A_{\rm P}^{t'}}{\tilde A_{\rm P}^{c'}-
\tilde A_{\rm P}^{t'}}\right].
\end{equation}
Consequently,  we obtain
\begin{equation}\label{xi-phiKS}
\xi_{\phi K_{\rm S}}^{(d)}=+e^{-i\phi_d}
\left[\frac{1+\lambda^2b e^{i\Theta}e^{-i\gamma}}{1+
\lambda^2b e^{i\Theta}e^{+i\gamma}}\right].
\end{equation}
The theoretical estimates of $b e^{i\Theta}$ 
suffer from large hadronic uncertainties. However, since this parameter enters 
(\ref{xi-phiKS}) in a doubly Cabibbo-suppressed way, we obtain the 
following expressions:\cite{RF-rev}
\begin{eqnarray}
{A}_{\rm CP}^{\rm dir}(B_d\to \phi K_{\rm S})&=&0+
{O}(\lambda^2)\\
{A}_{\rm CP}^{\rm mix}(B_d\to \phi K_{\rm S})&=&-\sin\phi_d
+{O}(\lambda^2),
\end{eqnarray}
where we made the plausible assumption that $b={O}(1)$. On the other 
hand, the mixing-induced CP asymmetry of 
$B_d\to J/\psi K_{\rm S}$ measures also $-\sin\phi_d$, as we saw in
(\ref{Amix-BdpsiKS}). We arrive therefore at the following 
relation:\cite{RF-rev}$^{\mbox{--}}$\cite{GLNQ}
\begin{equation}\label{Bd-phiKS-SM-rel}
{A}_{\rm CP}^{\rm mix}(B_d\to \phi K_{\rm S}) 
={A}_{\rm CP}^{\rm mix}(B_d\to J/\psi K_{\rm S}) + 
{O}(\lambda^2),
\end{equation}
which offers a very interesting test of the SM. Since $B_d\to \phi K_{\rm S}$ is 
governed by penguin processes in the SM, this decay may well be affected by 
NP. In fact, if we assume that NP arises generically in the TeV regime, it can be 
shown through field-theoretical estimates that the NP contributions to  
$b\to s\bar s s$ transitions may well lead to sizeable violations of
(\ref{Bd-phiKS-SM-rel}); in order to trace the origin of NP systematically,
a combined analysis of the neutral and the charged $B\to \phi K$ modes 
would be very useful.\cite{FM-BphiK}

\subsection{Experimental Status}
It is interesting to have a brief look at the time evolution of the $B$-factory 
data. At the LP '03 conference,\cite{LP03} the picture was as follows:
\begin{equation}
{A}_{\rm CP}^{\rm dir}(B_d\to \phi K_{\rm S})=
\left\{\begin{array}{ll}
-0.38\pm0.37\pm0.12 & \mbox{(BaBar)}\\
+0.15\pm0.29\pm0.07
& \mbox{(Belle)}
\end{array}\right.
\end{equation}
\begin{equation}
{A}_{\rm CP}^{\rm mix}(B_d\to \phi K_{\rm S})=
\left\{\begin{array}{ll}
-0.45\pm0.43\pm0.07 & \mbox{(BaBar)}\\
+0.96\pm0.50^{+0.11}_{-0.09} &
\mbox{(Belle).}
\end{array}\right.
\end{equation}
In the summer of last year, the following situation emerged at 
ICHEP '04:\cite{ICHEP04}
\begin{equation}
{A}_{\rm CP}^{\rm dir}(B_d\to \phi K_{\rm S})=
\left\{\begin{array}{ll}
+0.00 \pm 0.23 \pm 0.05 & \mbox{(BaBar\cite{Babar-BphiK-ICHEP04})}\\
-0.08 \pm 0.22 \pm 0.09
& \mbox{(Belle\cite{Belle-BphiK-ICHEP04})}
\end{array}\right.
\end{equation}
\begin{equation}
\underbrace{{A}_{\rm CP}^{\rm mix}(B_d\to 
\phi K_{\rm S})}_{\mbox{$\equiv-(\sin2\beta)_{\phi K_{\rm S}}$}}=
\left\{\begin{array}{ll}
-0.50 \pm 0.25_{-0.07}^{+0.04} 
&\mbox{(BaBar\cite{Babar-BphiK-ICHEP04})}\\
- 0.06 \pm 0.33 \pm 0.09 &
\mbox{(Belle\cite{Belle-BphiK-ICHEP04}).}
\end{array}\right.
\end{equation}

Because of 
$-(\sin2\beta)_{\psi K_{\rm S}}
\equiv{A}_{\rm CP}^{\rm mix}(B_d\to J/\psi K_{\rm S})=-0.725 \pm 0.037$,
the Belle data may indicate a violation of (\ref{Bd-phiKS-SM-rel}) through 
CP-violating NP contributions to $b\to s \bar s s$ transitions, which has 
already stimulated many speculations about NP effects in the decay 
$B_d\to\phi K_{\rm S}$.\cite{JN,BKK} However, the new Belle 
data moved towards 
the SM, and the BaBar data -- though also somewhat on the lower side -- are 
in accordance with the SM. Consequently, it seems too early to get too excited 
by the possibility of having a violation of the SM relation (\ref{Bd-phiKS-SM-rel}). 

It will be very interesting to observe how the $B$-factory data will evolve,
and to monitor also similar modes, such as $B_d\to \eta' K_{\rm S}$. However,
it is questionable to perform averages over many decays of this kind to argue
for NP in $b\to s$ penguin processes, as is frequently done in the literature.
The point is that we encounter different hadronic uncertainties in the SM, 
and that also NP is generally expected to affect these decays differently.

\begin{figure}[t] 
   \vspace*{-0.1truecm}
   \centering
   \includegraphics[width=10.0truecm]{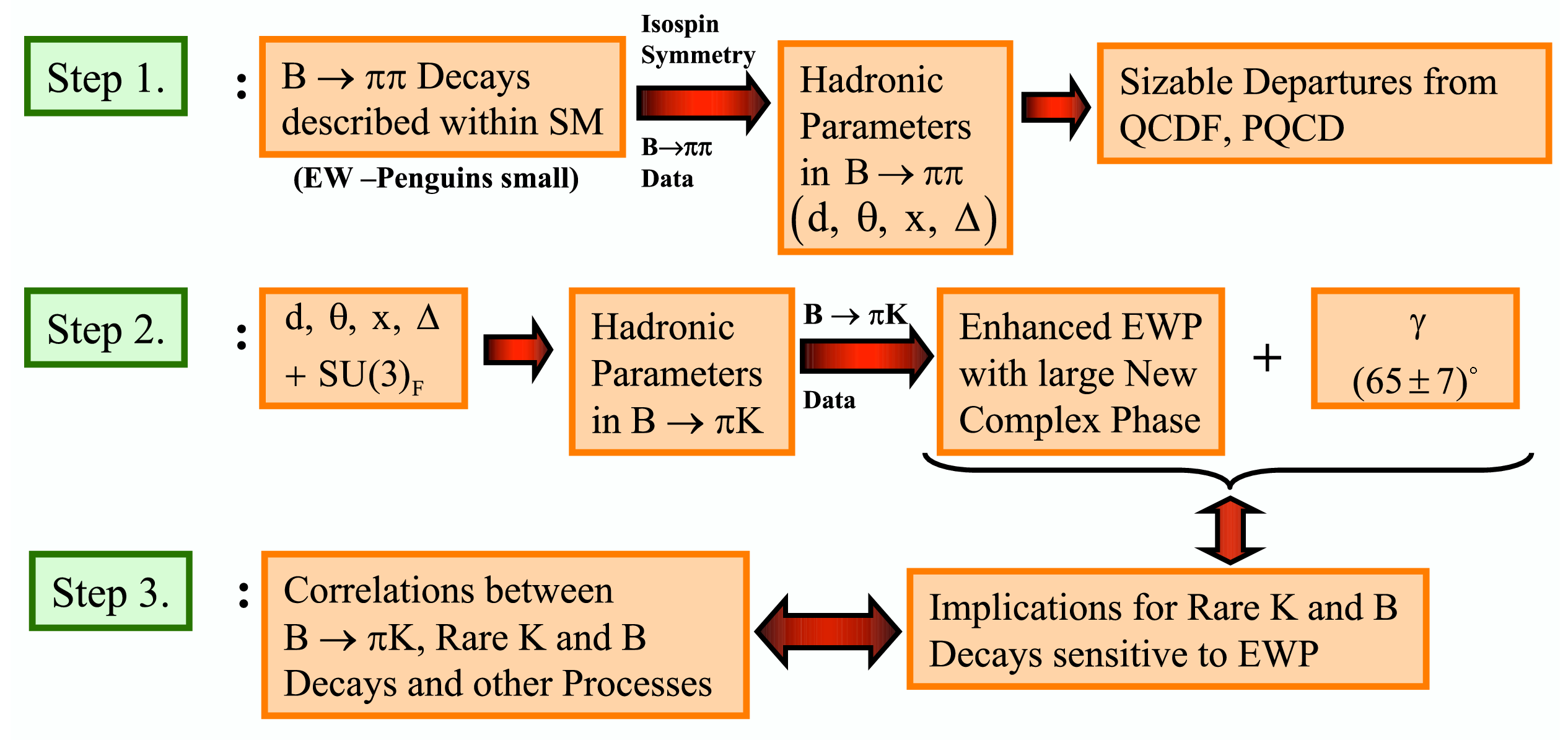} 
    \caption{The logical structure of a systematic strategy to analyze the 
    $B\to\pi\pi,\pi K$ puzzles and to explore their implications for rare $K$ and $B$ 
    decays.}\label{fig:chart}
\end{figure}

\section{The $B\to\pi\pi,\pi K$ Puzzles and their Implications for 
Rare $K$ and $B$ Decays}\label{sec:puzzles}
\subsection{Preliminaries}
For the search of NP signals and the exploration of their specific nature, 
it is crucial to exploit as much experimental information as possible, and 
to make also use of the interplay between CP-violating phenomena and 
rare decays.  As an example, let us discuss a strategy, which was recently proposed
to analyze puzzling patterns in the data for $B\to\pi\pi$ and $B\to\pi K$ decays, 
and to investigate their implications for rare $K$ and $B$ decays.\cite{BFRS} 
Its starting point is the SM, with 
\begin{equation}\label{BFRS-CKM}
\phi_d=(46.5^{+3.2}_{-3.0})^\circ~\mbox{(see  (\ref{phid-exp}))}, \quad \gamma=(65\pm7)^\circ~\mbox{(CKM fits),}
\end{equation}
and it consists of three interrelated steps, as illustrated in 
Fig.~\ref{fig:chart}:
\begin{itemize}
\item In step I, we perform
an isospin analysis of the currently available $B\to\pi\pi$ data, allowing us to 
extract a set of hadronic parameters characterizing the $B\to\pi\pi$ system, 
and to predict the CP-violating $B_d\to\pi^0\pi^0$
observables. We find large non-factorizable effects, but arrive at a picture
which is consistent with the SM. 
\item In step II, we use the hadronic 
$B\to\pi\pi$ parameters from step I to determine their $B\to\pi K$ counterparts 
with the help of the $SU(3)$ flavour symmetry, allowing us to predict the 
$B\to\pi K$ observables in the SM. We find agreement with the $B$-factory 
data in the case of those decays that are only marginally affected by EW penguins.
Moreover, we may extract $\gamma$, in excellent accordance with 
(\ref{BFRS-CKM}), and can perform a couple of other internal consistency 
checks, which also support our working assumptions. On the other hand, in the 
case of the $B\to\pi K$ decays with a significant impact of EW penguins, we obtain 
predictions which are {\it not} in agreement with the current data. This feature 
is a manifestation of the ``$B\to\pi K$ puzzle", which was already pointed out
in 2000,\cite{BF00} and received considerable attention in the recent literature (see, for instance, Refs.~81, 106--110).  
It can be resolved through NP in the EW penguin sector, which enhances the
corresponding contributions and introduces a new CP-violating phase.
\item In step III, we assume that NP enters in the EW penguin sector through $Z^0$ 
penguins, and explore the interplay of such a scenario with rare $K$ and $B$ 
decays. Interestingly, spectacular NP effects would arise in several processes, 
in particular in $K_{\rm L}\to\pi^0\nu\bar\nu$ and $B_{s,d}\to\mu^+\mu^-$ modes,
thereby leading to a specific pattern which can be {\it tested} experimentally. 
\end{itemize}
Let us now have a closer look at these three steps, where the numerical results 
refer to a recent update.\cite{BFRS-update}

\subsection{Step I: $B\to\pi\pi$}\label{ssec:Bpipi-puzzle}
\subsubsection{Input Observables}
The $B\to\pi\pi$ system offers three channels, 
$B^+\to\pi^+\pi^0$, $B^0_d\to\pi^+\pi^-$ and $B^0_d\to\pi^0\pi^0$,
as well as their CP conjugates. Consequently, two independent
ratios of the corresponding CP-averaged branching ratios are at
our disposal, which we may introduce as follows: 
\begin{eqnarray}
R_{+-}^{\pi\pi}&\equiv&
2\left[\frac{\mbox{BR}(B^\pm\to\pi^\pm\pi^0)}{\mbox{BR}
(B_d\to\pi^+\pi^-)}\right]\frac{\tau_{B^0_d}}{\tau_{B^+}} =
2.20\pm0.31\label{Rpm}\\
R_{00}^{\pi\pi}&\equiv&2\left[\frac{\mbox{BR}(B_d\to\pi^0\pi^0)}{\mbox{BR}
(B_d\to\pi^+\pi^-)}\right] = 0.67\pm0.14. \label{R00}
\end{eqnarray}
The branching ratios for $B_d\to\pi^+\pi^-$ and $B_d\to\pi^0\pi^0$ are found 
to be surprisingly small and large, respectively, whereas the one for
$B^\pm\to\pi^\pm\pi^0$ is in accordance with theoretical estimates. This feature
is reflected by the pattern of $R_{+-}^{\pi\pi}\sim 1.24$ and  
$R_{00}^{\pi\pi}\sim 0.07$ arising in QCDF.\cite{BeNe}
In addition to the CP-conserving observables in (\ref{Rpm}) and (\ref{R00}), 
we may also exploit the CP-violating observables of the $B_d\to\pi^+\pi^-$
decay:
\begin{eqnarray}
{A}_{\rm CP}^{\rm dir}(B_d\to\pi^+\pi^-)&=&
-0.37\pm 0.11\label{ACDdir} \\
{A}_{\rm CP}^{\rm mix}(B_d\to\pi^+\pi^-)&=&
+0.61\pm0.14.\label{ACPmix}
\end{eqnarray}
The experimental picture of these CP asymmetries
is not yet fully settled.\cite{HFAG} However, their theoretical 
interpretation discussed below yields constraints for the UT
in excellent agreement with the SM.

\subsubsection{Hadronic Parameters}
Using the isospin flavour symmetry of strong interactions, the observables in
(\ref{Rpm})--(\ref{ACPmix}) depend on two (complex) hadronic parameters,
$de^{i\theta}$ and $xe^{i\Delta}$, which describe --  sloppily speaking -- the 
ratio of penguin to colour-allowed tree amplitudes and the ratio of  
colour-suppressed to colour-allowed tree amplitudes, respectively. It is 
possible to extract these quantities {\it cleanly} and {\it unambiguously} from the
data:\footnote{EW penguin topologies have a tiny impact on the $B\to\pi\pi$ 
system, but are included in the numerical analysis.\cite{BFRS-update}}
\begin{equation}\label{Bpipi-par-det}
d=0.51^{+0.26}_{-0.20},\quad \theta=+(140^{+14}_{-18})^\circ,\quad x=
1.15^{+0.18}_{-0.16},\quad \Delta=-(59^{+19}_{-26})^\circ;
\end{equation}
a similar picture is also found by other authors.\cite{ALP}$^{\mbox{--}}$\cite{CGRS} 
In particular the impressive strong phases give an unambiguous signal for large deviations from
``factorization". In recent QCDF\cite{busa} and PQCD\cite{kesa} analyses, the 
following numbers were obtained:
\begin{equation}
\left.d\right|_{\rm QCDF}=0.29\pm0.09, \quad
\left.\theta\right|_{\rm QCDF}=-\left(171.4\pm14.3\right)^\circ, 
\end{equation}
\begin{equation}
\left.d\right|_{\rm PQCD}=0.23^{+0.07}_{-0.05}, \quad
+139^\circ < \left.\theta\right|_{\rm PQCD} < +148^\circ,
\end{equation}
which depart significantly from the experimental pattern in (\ref{Bpipi-par-det}).

\subsubsection{CP Violation in $B_d\to\pi^0\pi^0$}
Having the hadronic paramters of (\ref{Bpipi-par-det}) at hand, the CP-violating
asymmetries of the $B_d\to\pi^0\pi^0$ channel can be predicted:
\begin{eqnarray}
\left.{A}_{\rm CP}^{\rm dir}(B_d\to\pi^0\pi^0)
\right|_{\rm SM}&=&-0.28^{+0.37}_{-0.21}\label{Adir00-pred}\\
\left.{A}_{\rm CP}^{\rm mix}(B_d\to\pi^0\pi^0)\right|_{\rm SM}&=&
-0.63^{+0.45}_{-0.41},
\end{eqnarray}
thereby offering the exciting perspective of large CP violation in this
decay. The first results for the direct CP asymmetry were recently reported:
\begin{equation}
{A}_{\rm CP}^{\rm dir}(B_d\to \pi^0\pi^0)=
\left\{
\begin{array}{ll}
-(0.12\pm0.56\pm0.06) & \mbox{(BaBar\cite{babar-Adir00})}\\
-(0.43\pm0.51\,^{+0.17}_{-0.16}) & 
\mbox{(Belle\cite{belle-Adir00}),}
\end{array}
\right.
\end{equation}
and correspond to the average
of $A_{\rm CP}^{\rm dir}(B_d\to \pi^0\pi^0)=-(0.28\pm0.39)$,\cite{HFAG} 
which shows an encouraging agreement with (\ref{Adir00-pred}). In the 
future, more accurate input data will allow us to make much more stringent
predictions.

\subsection{Step II: $B\to\pi K$}\label{ssec:BpiK-puzzle}
\subsubsection{Ingredients of the $B\to\pi K$ Analysis}\label{ssec:BpiK-ingr}
In contrast to the $B\to\pi\pi$ modes, which originate from $b\to d$ processes,
we have to deal with $b\to s$ transitions in the case of the $B\to\pi K$ system. 
Consequently, these decay classes differ in their CKM structure and exhibit a 
very different dynamics. In particular, the $B\to\pi K$ decays are dominated by 
QCD penguins. Concerning the EW penguins, the $B\to\pi K$ decays can be 
divided as follows:
\begin{itemize}
\item $B^0_d\to\pi^-K^+$, $B^+\to\pi^+K^0$ (and CP conjugates): EW penguins
are colour-suppressed and expected to play a tiny r\^ole;
\item $B^+\to\pi^0K^+$, $B^0_d\to\pi^0K^0$ (and CP conjugates): EW penguins
are colour-allowed and have therefore a significant impact.
\end{itemize}
The starting point of our $B\to\pi K$ analysis are the hadronic $B\to\pi\pi$
parameters determined in Subsection~\ref{ssec:Bpipi-puzzle}, and the
values of $\phi_d$ and $\gamma$ in (\ref{BFRS-CKM}), which correspond to
the SM and are only insignificantly affected by EW penguins. We use then the 
following working hypothesis:
\begin{itemize}
\item[(i)] $SU(3)$ flavour symmetry of strong interactions;
\item[(ii)] neglect of penguin annihilation and exchange topologies.
\end{itemize}
It is important to stress that internal consistency checks of these 
assumptions can be performed, which are nicely satisfied by
the current data and do not indicate any anomalous behaviour. 
We may then determine the hadronic $B\to\pi K$ parameters 
through their $B\to\pi\pi$ counterparts, allowing us to predict 
the $B\to\pi K$ observables in the SM.

\subsubsection{Observables with a Tiny Impact of EW Penguins}\label{ssec:tiny-EWP}
Let us first have a look at the observables with a {\it tiny} impact
of EW penguins. Here the direct CP asymmetry in 
$B_d\to\pi^\mp K^\pm$ modes, which could be observed last summer, 
plays an important r\^ole. The average of the corresponding
BaBar and Belle results\cite{CP-dir-B} is given as follows:\cite{HFAG}
\begin{equation}
{A}_{\rm CP}^{\rm dir}(B_d\to\pi^\mp K^\pm)=
+0.113\pm 0.01,
\end{equation}
and establishes direct CP violation in the $B$-meson system.
The non-zero value of this CP asymmetry is generated through the interference 
between a QCD penguin and a colour-allowed tree amplitude, where the 
former dominates. In our strategy, we obtain the following prediction:
\begin{equation}
{A}_{\rm CP}^{\rm dir}(B_d\to\pi^\mp K^\pm)=+0.127^{+0.102}_{-0.066},
\end{equation}
which agrees nicely with the experimental value. Moreover, assumptions (i) 
and (ii) listed in Subsection~\ref{ssec:BpiK-ingr} imply the following relation:
\begin{equation}
H\propto\underbrace{\left(\frac{f_K}{f_\pi}\right)^2\left[\frac{\mbox{BR}
(B_d\to\pi^+\pi^-)}{\mbox{BR}(B_d\to\pi^\mp K^\pm)}
\right]}_{\mbox{0.38$\pm$0.04}}=
\underbrace{-\left[\frac{{A}_{\rm CP}^{\rm dir}(B_d\to\pi^\mp 
K^\pm)}{{A}_{\rm CP}^{\rm dir}(B_d\to\pi^+\pi^-)}
\right]}_{\mbox{0.31$\pm$0.11}},
\end{equation}
where we have also indicated the experimental values, which give us
further confidence into our working assumptions. Moreover, since we
may write
\begin{equation}
H=G_3(d,\theta;\gamma),
\end{equation}
the $B_d\to\pi^\mp K^\pm$ data allow us to convert the CP asymmetries
\begin{eqnarray}
{A}_{\rm CP}^{\rm dir}(B_d\to\pi^+\pi^-)&=&
G_1(d,\theta;\gamma) \\
{A}_{\rm CP}^{\rm mix}(B_d\to\pi^+\pi^-)&=&
G_2(d,\theta;\gamma,\phi_d)
\end{eqnarray}
into a value of $\gamma$.\cite{FlMa,RF-BsKKBdpipi} The corresponding result
is shown as the quadrangle in Fig.~\ref{fig:UT-fit}, which is in excellent
agreement with all the other UT constraints.

On the other hand, a moderate numerical discrepancy arises for the ratio $R$ of 
the CP-averaged $B_d\to\pi^\mp K^\pm$, $B^\pm\to\pi^\pm K$
branching ratios.\cite{FM} This feature suggests a sizeable impact of a
hadronic parameter $\rho_{\rm c}e^{i\theta_{\rm c}}$, which enters the 
most general parametrization of the $B^+\to\pi^+K^0$ amplitude.\cite{defan,neubert}
It can be constrained through the direct CP asymmetry of the 
decay $B^\pm\to\pi^\pm K$ and the emerging $B^\pm\to K^\pm K$ signal, and
actually shifts the predicted value of $R$ towards the data.\cite{BFRS-update} 
Consequently, no discrepancies with the SM arise in this sector 
of the $B\to\pi K$ system.

\subsubsection{Observables with a Sizeable Impact of EW Penguins}
Let us now turn to those observables that are significantly affected by
EW penguins. The key quantities are the following ratios:\cite{BF98}
\begin{equation}
R_{\rm c}\equiv2\left[\frac{\mbox{BR}(B^+\to\pi^0K^+)+
\mbox{BR}(B^-\to\pi^0K^-)}{\mbox{BR}(B^+\to\pi^+ K^0)+
\mbox{BR}(B^-\to\pi^- \bar K^0)}\right]\stackrel{\rm Exp}{=}1.00\pm0.08
\end{equation}
\begin{equation}
R_{\rm n}\equiv\frac{1}{2}\left[
\frac{\mbox{BR}(B_d^0\to\pi^- K^+)+
\mbox{BR}(\bar B_d^0\to\pi^+ 
K^-)}{\mbox{BR}(B_d^0\to\pi^0K^0)+
\mbox{BR}(\bar B_d^0\to\pi^0\bar K^0)}\right]
\stackrel{\rm Exp}{=}0.79\pm0.08,
\end{equation}
where the EW penguin contributions enter in colour-allowed form
through the decays with $\pi^0$-mesons in the final states. Theoretically,
the EW penguin effects are described by the following parameters:
\begin{equation}
q\stackrel{\rm SM}{=}0.69, \quad
\phi\stackrel{\rm SM}{=}0^\circ.
\end{equation}
Here $q$, which can be calculated in the SM with the help of the $SU(3)$ 
flavour symmetry,\cite{NR} measures the ``strength'' of the EW penguins 
with respect to the tree contributions, and $\phi$ is a CP-violating weak phase 
with an origin lying beyond the SM. EW penguin topologies offer an interesting 
avenue for NP to manifest itself, as is already known for 
several years.\cite{FM-BpiK-NP,trojan}

\begin{figure}[t] 
   \centering
   \vspace*{-0.2truecm}
   \includegraphics[width=8.2truecm]{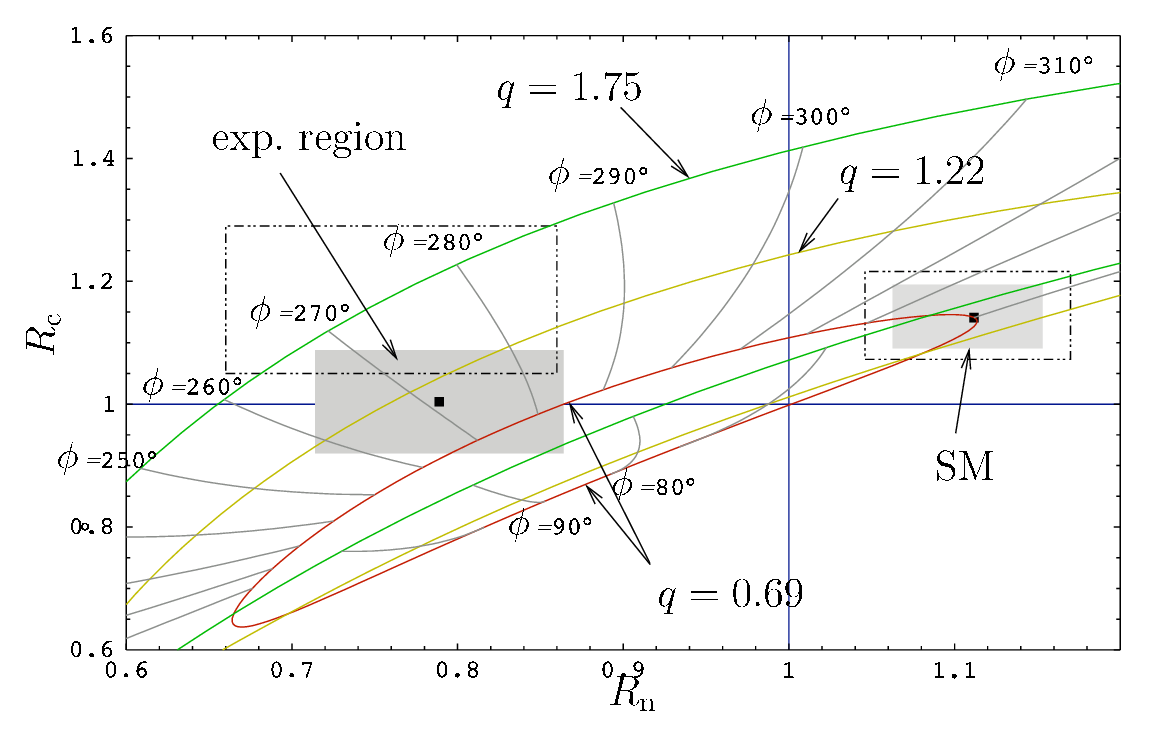} 
   \vspace*{-0.2truecm}
   \caption{The situation in the $R_{\rm n}$--$R_{\rm c}$ plane,
   as discussed in the text.}
   \label{fig:RnRc-update}
\end{figure}

In Fig.~\ref{fig:RnRc-update}, we have shown the current situation in the 
$R_{\rm n}$--$R_{\rm c}$ plane: the experimental ranges and those 
predicted in the SM are indicated in grey, and the dashed lines serve as a 
reminder of the corresponding ranges in Ref.~42; 
the central values for the SM prediction have hardly moved, while their
uncertainties have been reduced a bit. Moreover, we show contours 
for values of $q=0.69$, $q=1.22$ and $q=1.75$, with
$\phi \in [0^\circ,360^\circ]$. We observe that we arrive no longer
at a nice agreement between our SM predictions and the experimental
values. However, as becomes obvious from the contours
in Fig.~\ref{fig:RnRc-update}, this discrepancy can be resolved if we allow for
NP in the EW penguin sector, i.e.\ keep $q$ and $\phi$ as free
parameters. Following these lines, the successful picture described above 
would not be disturbed, and we obtain full agreement between the theoretical 
values of $R_{\rm n, c}$ and the data. The corresponding values of $q$
and $\phi$ are given as follows:
\begin{equation}\label{q-phi-det}
q = 1.08\,^{+0.81}_{-0.73}, 
\quad 
\phi = -(88.8^{+13.7}_{-19.0})^\circ,
\end{equation}
where in particular the large CP-violating phase would be a
striking signal of NP. These parameters allow us then to 
predict also the CP-violating observables of the $B^\pm\to\pi^0 K^\pm$
and $B_d\to\pi^0K_{\rm S}$ decays,\cite{BFRS-update} which should
provide useful tests of this scenario in the future.
Particularly promising in this respect are rare $K$ and $B$ decays.

\subsection{Step III: Rare $K$ and $B$ Decays}\label{ssec:rareKB}
In order to explore the implications for rare $K$ and $B$ decays, we
assume that NP enters the EW penguin sector through enhanced 
$Z^0$ penguins with a new CP-violating phase. This scenario, which 
belongs to class C introduced in Subsection~\ref{ssec:NP-class}, was already
considered in the literature, where model-independent analyses and 
studies within SUSY were presented.\cite{Z-pen-analyses,BuHi} 
In our strategy, we determine the short-distance function $C$ characterizing 
the $Z^0$ penguins through the $B\to\pi K$ data. Performing a 
renormalization-group analysis,\cite{BFRS-I} we obtain
\begin{equation}\label{RG}
C(\bar q)= 2.35~ \bar q e^{i\phi} -0.82 \quad\mbox{with}\quad 
\bar q= q \left[\frac{|V_{ub}/V_{cb}|}{0.086}\right].
\end{equation}
If we evaluate then the relevant box-diagram contributions within the SM 
and use (\ref{RG}), we can calculate the short-distance functions
\begin{equation}\label{X-C-rel}
X=2.35~ \bar q e^{i\phi} -0.09 \quad \mbox{and} \quad 
Y=2.35~ \bar q e^{i\phi} -0.64,
\end{equation}
which govern the rare $K$, $B$ decays with $\nu\bar\nu$ and $\ell^+\ell^-$ 
in the final states, respectively. In the SM, we have $C=0.79$, $X=1.53$
and $Y=0.98$, with {\it vanishing} CP-violating phases.

\begin{table}[ph]
\tbl{Comparison of the predicted values of
$R_{\rm c}$ and $R_{\rm n}$ taking the constraints from rare decays 
into account with the evolution of the data, as discussed in the text.}
{\footnotesize
\begin{tabular}{@{}cccc@{}}
\hline
{} &{} &{} &{} \\[-1.5ex]
{} & ``Old" data & Prediction with RDs & ``New" data\\[1ex]
\hline
{} &{} &{} &{}\\[-1.5ex]
$R_{\rm c}$ & $1.17\pm0.12$ & $1.00^{+0.12}_{-0.08}$ & $1.00\pm0.08$ \\[1ex]
$R_{\rm n}$ & $0.76\pm0.10$ & $0.82^{+0.12}_{-0.11}$ & $0.79\pm0.08$ \\[1ex]
\hline
\end{tabular}\label{tab} }
\end{table}

If we impose constraints from the data for rare decays, in particular those on
$|Y|$ following from $B\to X_s\mu^+\mu^-$, the following picture arises:
\begin{equation}\label{q-phi-RD}
\bar q= 0.92^{+0.07}_{-0.05},\quad
\phi=-(85^{+11}_{-14})^\circ.
\end{equation}
In Table~\ref{tab}, we compare the corresponding predictions of
$R_{\rm c}$ and $R_{\rm n}$ with the ``old" data, which were 
available when these predictions were made,\cite{BFRS} and 
the ``new" data, which emerged at the ICHEP '04 
conference.\cite{ICHEP04} We observe that the data 
have moved accordingly.

The values in (\ref{q-phi-RD}) are compatible with all the current data on
rare decays, and are in accordance with the new $B\to\pi K$ data. 
However, we may still encounter significant deviations from the SM
expectations for certain rare decays, with a set of predictions that is characteristic
for our specific NP scenario, thereby allowing an experimental {\it test}
of this picture. The most spectacular effects are the following ones:
\begin{itemize}
\item $\mbox{BR}(K_{\rm L}\to\pi^0\nu\bar\nu)$ is enhanced
by a factor of ${O}(10)$, which brings it close to the Grossman--Nir 
bound,\cite{GN-bound} whereas  
$\mbox{BR}(K^+\to\pi^+\nu\bar\nu)$ remains essentially unchanged.
Consequently, we would also have a strong violation of the following MFV 
relation:\cite{BuBu}
\begin{equation}
\underbrace{(\sin 2 \beta)_{\pi \nu\bar\nu}}_{\mbox{$-(0.69^{+0.23}_{-0.41})$}}=
\underbrace{(\sin 2 \beta)_{\psi K_{\rm S}}}_{\mbox{$+(0.725\pm0.037)$}},
\end{equation}
where we have indicated the corresponding numerical values.

\item The decay $K_{\rm L}\to\pi^0e^+e^-$ would now be governed by direct 
CP violation, and its branching ratio would be enhanced by a factor of 
${O}(3)$. The interesting implicatios for $K_{\rm L}\to\pi^0\mu^+\mu^-$ were
discussed in a recent paper.\cite{ISU}

\item In the case of $B_d\to K^*\mu^+\mu^-$, an integrated forward--backward 
CP asymmetry\cite{BuHi} can be very large, whereas it vanishes in the SM. The 
corresponding NP effects for the lepton polarization asymmetries of 
$B\to X_s\ell^+\ell^-$ decays were recently studied.\cite{CCG}

\item The branching ratios for $B\to X_{s,d}\nu\bar\nu$ and 
$B_{s,d}\to \mu^+\mu^-$ decays would be enhanced by factors of $2$ and $5$, 
respectively, whereas the impact on $K_{\rm L}\to \mu^+\mu^-$ is 
rather moderate.
\end{itemize}
If future, more accurate, $B\to\pi\pi,\pi K$ data will not significantly modify
the currently observed patterns in these decays, the scenario
of enhanced $Z^0$ penguins with a large CP-violating NP phase $\phi$
will remain an attractive scenario for physics beyond the SM. It will then
be very interesting to confront the corresponding predictions for the rare
$K$ and $B$ decays listed above with experimental results.

\section{Conclusions and Outlook}\label{sec:concl}
Flavour physics offers interesting strategies to explore the SM and 
to search for signals of NP. In the $B$-meson system, data from the 
$e^+e^-$ $B$ factories agree on the one hand remarkably well with 
picture of the Kobayashi--Maskawa mechanism, where the accordance 
between the measurement of $\sin 2\beta$ through $B_d\to J/\psi K_{\rm S}$ 
decays and the CKM fits is the most important example. On the other 
hand,  there are also hints for discrepancies with the SM, and it will be 
very interesting to monitor these effects in the future. 

Despite this impressive progress, there are still regions of the 
$B$-physics ``landscape"  left that are unexplored. 
For instance, $b\to d$ penguin processes are now close to enter 
the stage, since lower bounds for the corresponding branching 
ratios that can be derived in the SM are found to be close to the current 
experimental upper limits.\cite{FR-bounds} In fact, the BaBar collaboration 
has already reported the first signals for the $B_d\to K^0\bar K^0$ channel, 
in accordance with these bounds.\cite{BdKK-babar} The lower SM bounds for
other non-leptonic decays of this kind and for $B\to\rho\gamma$ transitions
suggest that these modes should also be observed soon. For the more 
distant future, decays such as $B\to \rho\ell^+\ell^-$ decays are left. Since
the various $b\to d$ penguin modes are governed by different operators, 
they may be affected differently by NP. Moreover, as we have 
emphasized throughout these lectures, also the $B_s$-meson system is still 
essentially unexplored, and offers a very promising physics potential
for the search of NP, which should be fully exploited -- after first steps 
at run II of the Tevatron -- at the LHC, in particular by LHCb. 

These studies can nicely be complemented through the kaon system, which 
governed the stage of CP violation for more than 35 years. The future lies now 
on rare decays, in particular on the $K\to\pi\nu\bar\nu$ modes, and any effort 
should be made to measure these very challenging -- but also very rewarding -- 
decays; this is the goal of the NA48 (CERN), E391(a) (KEK/J-PARC) and
KOPIO (BNL) experiments. 

In addition to these electrifying aspects, flavour physics offers many more
exciting topics, which we could unfortunately not cover here. Important
examples are the $D$-meson system, electric dipole moments, and
the search for flavour-violating charged lepton decays.

For the search of NP and the exploration of its nature, it is important to 
keep an eye on all of these processes and to aim for the {\it whole} picture. 
In particular correlations between various $K$ and $B$ decays play 
an outstanding r\^ole in this context, as we have illustrated through the 
discussion of the $B\to\pi K$ puzzle. 
A fruitful interplay between flavour physics and the direct NP searches by 
ATLAS and CMS at the LHC is also expected, and will soon be explored in 
much more detail.\cite{workshop} I have no doubt that an exciting future is 
ahead of us!

\section*{Acknowledgments}
I would like to thank the organizers for inviting me to this wonderful
Winter Institute in such a spectacular environment, and would also
like to thank the participants for their stimulating interest in my lectures.

\end{document}